\def\thetac{\theta_C}

\def\dnuup{\partial^{\nu}}

\def\dsone{|\Delta S|=1}

\def\smalllinestretch{\renewcommand{\baselinestretch}{0.7}}
\def\largelinestretch{\renewcommand{\baselinestretch}{1.0}}
 \def\tr {\, \mbox{tr} \,}

 \def\im {\, \mbox{Im} \,}
 \def\re {\, \mbox{Re} \,}
 \def\hc {   \mbox{h.c.} }
 \def\kb{\overline{K}^0}

 \def\L{{\cal L}}
 \def\M{{\cal M}}
 \def\Mb{\overline{\cal M}}
 \def\O{{\cal O}}
 \def\D{{\cal D}}
 \def\Db{\overline{\cal D}}
 \def\Z{{\cal Z}}

 \def\qq{<\!\!\bar{q} q\!\!>}

 \def\Zz{\Z_9}
 \def\Zzz{\Z_{10}}
 \def\up{U^+}
 \def\fj{\varphi}
 \def\Fi{\Phi}

 \def\dmu{\partial_{\mu}}
 \def\dnu{\partial_{\nu}}

 \def\dii{\partial^2}

 \def\dmuup{\partial^{\mu}}
 \def\dnuup{\partial^{\nu}}

 \def\Dmu{\D_{\mu}}
 
 \def\Dal{\D_{\alpha}}
 
 \def\Dii{\D^2}

 \def\Dmuup{\D^{\mu}}
 \def\Dnuup{\D^{\nu}}
 
 \def\Dalup{\D^{\alpha}}

 \def\Dbmu{\Db_{\mu}}
 \def\Dbnu{\Db_{\nu}}
 \def\Dbal{\Db_{\alpha}}
 
 \def\Dbii{\Db^2}

 \def\Dbmuup{\Db^{\mu}}
 \def\Dbnuup{\Db^{\nu}}
 \def\Dbalup{\Db^{\alpha}}

 \def\Fpmunu{F^{(+)}_{\mu \nu}}
 \def\Fmmunu{F^{(-)}_{\mu \nu}}

 \def\diag{\mbox{diag}}
 \def\mev{\,\mbox{MeV}}
 \def\gev{\,\mbox{GeV}}
 \def\fm{\,\mbox{fm}}

 \def\gt{\widetilde{G}}

\newcommand{\su}[1]{\mbox{\small\it #1}}
\documentstyle{article}
\def\largelinestretch{\renewcommand{\baselinestretch}{1.2}}
\largelinestretch\small\normalsize
\title{\huge Radiative Kaon Decays and $CP$ Violation}
 \author{
A.A.Bel'kov${}^1$,
A.V.Lanyov${}^1$,
A.Schaale${}^2$
\\
\\
\small
${}^1$
        Particle Physics Laboratory, Joint Institute for Nuclear Research,
\hfill\\
\small
        Head Post Office, P.O. BOX 79, 101000 Moscow, Russia
\hfill\\
\small
${}^2$
        DESY-Institute for High Energy Physics,
\hfill\\
\small
        Platanenallee 6, O-1615 Zeuthen, Germany
}

\begin{document}
\largelinestretch\normalsize
\begin{titlepage}
\maketitle
\begin{abstract}
    The amplitudes of $K \rightarrow \pi \gamma^* \rightarrow \pi e^+e^-$
and $K \rightarrow \pi \pi \gamma$ decays have been calculated within
chiral Lagrangian approach including higher-order derivative terms and
meson loops.
    The selfconsistency of the simultaneous description of the
experimental data on the nonleptonic and radiative kaon decays have been
demonstrated.
    We estimate the effects of ``indirect'' and ``direct'' $CP$-violation
in $K^0_L \rightarrow \pi^0 e^+e^-$ decays and discuss
$CP$-violating charge asymmetries in
$K^{\pm} \rightarrow \pi^{\pm} \pi^0 \gamma$ decays.
%
%

\end{abstract}
\end{titlepage}

   The radiative kaon decays constantly attract the attention both of
theoreticians and experimentalists as a possible alternative sources
of the information on $CP$-violation in addition to
$K^0 \rightarrow 2\pi$ decays.
   The question, if $CP$-noninvariant interactions do indeed give rise
to the experimentally observable $CP$-noninvariant effects, is closely
connected with the verification of effective chiral Lagrangians which
are widely used now for the theoretical calculations of
$CP$-conserving and $CP$-violating parts of kaon decay amplitudes.
   In this paper the radiative one-photon
$K \rightarrow \pi \gamma^* \rightarrow \pi e^+e^-$ and
$K \rightarrow \pi \pi \gamma$ decays are considered within the approach
based on the bosonization of effective strong, weak, and
electromagnetic-weak quark interactions.
   The main purpose of this paper is to continue the investigation of
kaon decays with the bosonized effective meson Lagrangians which were
used in our previous paper \cite{DESY92-106} for the
description of nonleptonic kaon decays and for justification of the
enhancement of $CP$-violating charge asymmetries in $K \rightarrow 3\pi$
decays.
   To check the possibility of selfconsistent description of
nonleptonic and radiative kaon decays we used in this paper the values
of Wilson parameters fixed in \cite{DESY92-106} from the
analysis of the experimental data on $K \rightarrow 2\pi$,
$K \rightarrow 3\pi$ decays.

   First, we state the main definitions and assumptions, and
display all parameters entering the calculations.
   Then, in Section 2, the bosonized weak and electromagnetic-weak
currents will be applied for the description of form factors of
semileptonic and radiative semileptonic decays
$K, \pi \rightarrow \pi l \nu$, $K, \pi \rightarrow l \nu \gamma$,
$K, \pi \rightarrow \pi l \nu \gamma$.
   In Section 3 from chiral Lagrangians with higher-order derivative
terms and additional higher-order $s$-quark mass corrections,
including meson loops and electromagnetic penguins, the form-factors
of $K \rightarrow \pi \gamma^*$ transitions for charged and neutral
kaons are calculated.
   The ``indirect'' and ``direct'' $CP$-violation in
$K^0_L \rightarrow \pi^0 \gamma^* \rightarrow \pi^0 e^+e^-$ decays are
also estimated in the dependence on top quark mass.
   The decays $K \rightarrow \pi \pi \gamma$ are considered in Section 4
where we calculate the inner bremsstrahlung and direct emission parts
of amplitudes and estimate their contributions to the branching ratios
and spectrums of the center-of-mass $\gamma$ energy for
$K^+ \rightarrow \pi^+ \pi^0 \gamma$ and
$K^0_L \rightarrow \pi^+ \pi^- \gamma$ channels.
   The $CP$-violating charge asymmetries in
$K^{\pm} \rightarrow \pi^{\pm} \pi^0 \gamma$ decays are estimated and
the suppression of these effects is also discussed.

%
%
%
%
\section*{1. Lagrangians and currents}
%

  The effective Lagrangian describing nonleptonic weak interactions
with strangeness change $\dsone{}$ is given on the quark
level by \cite{vzsh,gilman-wise,bijnens-wise}:
 \begin{equation}
 \L^{\su{nl}}_{\su{w}} = \gt \sum_{i=1}^8 c_i \, {\cal O}_i \; + \hc
 \label{weak-lagr}
 \end{equation}
   Here $ \gt = \sqrt2 \, G_F \, \sin \thetac \,\cos \thetac$ is the
weak coupling constant;
   $c_i$ are Wilson coefficient functions which may be calculated in
the QCD leading-log approximation, depending then explicitly on the
renormalization scale $\mu$.
   The terms ${\cal O}_i$ are the four-quark operators
consisting of products of left- and/or right-handed quark currents:
 \begin{eqnarray}
\O_1 &=&   \bar{u}_L \gamma_\mu u_L \; \bar{d}_L \gamma^\mu s_L
       -   \bar{d}_L \gamma_\mu u_L \; \bar{u}_L \gamma^\mu s_L,
\nonumber\\
\O_2 &=&   \bar{u}_L \gamma_\mu u_L \; \bar{d}_L \gamma^\mu s_L
       +   \bar{d}_L \gamma_\mu u_L \; \bar{u}_L \gamma^\mu s_L
       + 2 \bar{d}_L \gamma_\mu d_L \; \bar{d}_L \gamma^\mu s_L
       + 2 \bar{s}_L \gamma_\mu s_L \; \bar{d}_L \gamma^\mu s_L,
\nonumber\\
\O_3 &=&   \bar{u}_L \gamma_\mu u_L \; \bar{d}_L \gamma^\mu s_L
       +   \bar{d}_L \gamma_\mu u_L \; \bar{u}_L \gamma^\mu s_L
       + 2 \bar{d}_L \gamma_\mu d_L \; \bar{d}_L \gamma^\mu s_L
       - 3 \bar{s}_L \gamma_\mu s_L \; \bar{d}_L \gamma^\mu s_L,
\nonumber\\
\O_4 &=&   \bar{u}_L \gamma_\mu u_L \; \bar{d}_L \gamma^\mu s_L
       +   \bar{d}_L \gamma_\mu u_L \; \bar{u}_L \gamma^\mu s_L
       -   \bar{d}_L \gamma_\mu d_L \; \bar{d}_L \gamma^\mu s_L,
\nonumber\\
\O_5 &=& \bar{d}_L \gamma_\mu \lambda^a_c s_L
  \left( \sum_{q=u,d,s} \bar{q}_R \, \gamma^\mu \, \lambda^a_c \, q_R \right),
\quad
\O_6 = \bar{d}_L \gamma_\mu s_L
  \left( \sum_{q=u,d,s} \bar{q}_R \, \gamma^\mu \, q_R \right),
\nonumber\\
\O_7 &=& \bar{d}_L \gamma_\mu s_L
  \left( \sum_{q=u,d,s} \bar{q}_R \, \gamma^\mu \, Q \, q_R \right),
\quad
\O_8 = \bar{d}_L \gamma_\mu \lambda^a_c s_L
  \left( \sum_{q=u,d,s} \bar{q}_R \, \gamma^\mu \, \lambda^a_c \, Q \, q_R
\right).
 \label{four-quark}
 \end{eqnarray}
Here $q_{L,R} = \frac12 \, (1\mp\gamma_5) q$; $\lambda^a_c$ are
the generators of the $SU(N_c)$ color group; $Q$ is the matrix of
electric quark charges.
The operators ${\cal O}_{5,6}$ containing right-handed
currents are generated by gluonic penguin diagrams and the analogous
operators ${\cal O}_{7,8}$ arise from electromagnetic
penguin diagrams.

   The bosonized version of the effective Lagrangian (\ref{weak-lagr})
can be expressed in the form \cite{91-08}:
 \def\jim{(J^1_{L\mu} - i J^2_{L\mu})}
 \def\rim{(J^1_R - i J^2_R)}
 \def\jiii{(J^3_{L\mu} + \frac1{\sqrt3} \, J^8_{L\mu})}
 \def\riii{(J^3_R - \frac1{\sqrt3} \, J^8_R - \sqrt\frac23 \, J^0_R)}
 \def\jivp{(J^4_{L\mu} + i J^5_{L\mu})}
 \def\livp{(J^4_L + i J^5_L)}
 \def\jvip{(J^6_{L\mu} + i J^7_{L\mu})}
 \def\lvip{(J^6_L + i J^7_L)}
 \def\rvip{(J^6_R + i J^7_R)}
 \begin{eqnarray}
 {\cal L}^{\su{nl}}_{\su{eff}} &=& \gt \bigg\{
  (-\xi_1 + \xi_2 + \xi_3) \bigg[ \jim \jivp - \jiii \jvip \bigg]
\nonumber\\
&&+ (\xi_1 + 5\,\xi_2) \sqrt\frac23 \, J^0_{L\mu} \jvip
+ {10 \over \sqrt3} \, \xi_3 \, J^8_{L\mu} \jvip
\nonumber\\
&&+ \xi_4 \bigg[ \jim \jivp + 2 \, J^3_{L\mu} \jvip \bigg]
\nonumber\\
&&- 4 \, \xi_5 \bigg[ \rim \livp - \riii \lvip
\nonumber\\
&&- \sqrt\frac23 \, \rvip (\sqrt2 J^8_L - J^0_L)
           \bigg]
\nonumber\\
&&+ \xi_6 \, \sqrt\frac32 \, \jivp J^0_R
+ 6 \, \xi_7 \, \jvip (J^3_{R\mu} + \frac1{\sqrt3} \, J^8_{R\mu})
\nonumber\\
&&- 16 \, \xi_8 \, \bigg[ \rim \livp + \frac12 \, \riii \lvip
\nonumber\\
&&\qquad + \frac1{\sqrt6} \, \rvip (\sqrt2 \, J^8_L - J^0_L)
            \bigg]
 \bigg\} + \hc
 \label{weak-bosl}
 \end{eqnarray}
Here $J^a_{L/R \, \mu}$ and $J^a_{L/R}$ are bosonized $(V\mp A)$
and $(S\mp P)$ meson currents corresponding to the quark
currents
$\bar q \gamma_\mu \frac14 (1\mp \gamma^5) \lambda^a q$
and
$\bar q \frac14 (1\mp \gamma^5) \lambda^a q$,
respectively ($\lambda^a$ are the generators of the $U(3)_F$ flavor group);
\begin{eqnarray}
&&\xi_1 = c_1 \left( 1 - {1 \over N_c} \right), \qquad
\xi_{2,3,4} = c_{2,3,4} \left( 1 + {1 \over N_c} \right),
\nonumber\\
&&\xi_{5,8} = c_{5,8} + {1 \over 2 N_c} c_{6,7}, \qquad
\xi_{6,7} = c_{6,7} - {2 \over N_c} c_{5,8},
\end{eqnarray}
where the color factor ${1/N_c}$ originates from the Fierz-transformed
contribution to the nonleptonic weak effective chiral Lagrangian
\cite{91-08}.

   Using the bosonization procedure described in \cite{91-08} the meson
currents in eq.(\ref{weak-bosl}) can be derived from the
various parts of the effective Lagrangian of strong interactions
arising from the divergent and finite parts of quark determinant in
QCD-motivated chiral quark model \cite{91-08,ebert-reinhardt}:
\begin{eqnarray}
 \L^{(div)} &=& {F_0^2 \over 4} \, \tr(\Dmu U \, \Dbmuup \up)
 + {F_0^2 \over 4} \, \tr ( M U + \hc ) ,
\label{4a}
\\
 \L^{(fin)}_1 &=& {N_c \over 32\pi^2} \, \tr
      \left[ \frac16 \, \big(\Dmu U \, \Dbnu \up \big)^2
             - \M \Dmu U \Dbmuup \up -\Mb \, \Dbmuup \up \Dmu U
      \right] ,
\label{4b}
\\
 \L^{(fin)}_2 &=& \frac{N_c}{32 \pi^2} \, tr \,
\bigg \{ \Zz \frac{2}{3} \Big( \Fpmunu \Dbmuup \up \Dnuup U
+ \Fmmunu \Dmuup U \Dbnuup \up \Big)
\nonumber
\\
&& + \frac{1}{3} \Zzz F^{(+) \mu \nu} \up \Fmmunu U
- \frac{5}{6m^2} \Big( \Mb ( \Fpmunu )^2 + \M ( \Fmmunu )^2 \Big)
\nonumber
\\
&& + \frac{1}{6m^2} \bigg[ \Dalup \Dmuup U \Big(
\Fpmunu \Dbal \Dbnuup \up - \Dal \Dnuup \up \Fmmunu \Big)
\nonumber
\\
&& - \Fmmunu \Big( \Dii \Dmuup U \Dbnuup \up
                  - \Dnuup U \Dbii \Dbmuup \up \Big)
   - \Fpmunu \Big( \Dbii \Dbmuup \up \Dnuup U
                  - \Dbnuup \up \Dii \Dmuup U \Big)
\nonumber
\\
&& + \frac{1}{10} \Big( 4 \Fpmunu \Dbalup \up F^{(-) \mu \nu } \Dal U
+ 3( \Fpmunu )^2 \Dbalup \up \Dal U + 3( \Fmmunu )^2 \Dalup U \Dbal \up
                  \Big)
\nonumber
\\
&& - 2 \Dbmuup \up \Big( \Dnuup U \{ \Fpmunu , \Mb \}
- \{ \Fmmunu , \M  \} \Dnuup U
+ \Mb \Dnuup U \Fpmunu - \Fmmunu \Dnuup U \M \Big)
\nonumber
\\
&& + 2 \bigg( F^{(-)}_{\mu \alpha} F^{(-) \alpha}_{\nu}
             \Big( \Dmuup U \Dbnuup \up - \Dnuup U \Dbmuup \up \Big)
            + F^{(+)}_{\mu \alpha} F^{(+) \alpha}_{\nu}
             \Big( \Dbmuup \up \Dnuup U - \Dbnuup \up \Dmuup U \Big)
\nonumber
\\
&&          - \Fmmunu \Dal U F^{(+) \nu \alpha} \Dbmuup \up
            - F^{(-) \nu \alpha} \Dal U \Fpmunu \Dbmuup \up
       \bigg)
             \bigg ] \bigg \} .
\label{4c}
\end{eqnarray}
   Here $U = \Omega^2$,
$\Omega = \exp \left\{ {i \Phi \over \sqrt2 \, F_0} \right\}$,
   where
$\Fi = \frac1{\sqrt3}\fj_0 + \frac1{\sqrt2} \sum_{a=1}^8 \lambda_a\fj_a$
   is the matrix of pseudoscalar meson fields $\fj_a$,
and $F_0$ being the (bare) decay constant of the $\pi \to \mu\nu$ decay.
$F^{( \pm )}_{ \mu \nu }$ are external field strength tensors defined as
$$
F^{( \pm )}_{ \mu \nu } = F^{V}_{ \mu \nu } \pm F^{A}_{ \mu \nu },
$$
$$
F^{V}_{ \mu \nu } = \dmu V_{\nu} - \dnu V_{\mu} + [V_{\mu} ,V_{\nu} ]
                    + [A_{\mu} ,A_{\nu} ],
\qquad
F^{A}_{ \mu \nu } = \dmu A_{\nu} - \dnu A_{\mu} + [V_{\mu} ,A_{\nu} ]
                    + [A_{\mu} ,V_{\nu} ],
$$
   and the covariant derivatives are defined as
\begin{eqnarray}
\Dmu U = \dmu U + [V_{\mu} , U ] - \{ A_{\mu} , U\} , \qquad
\Dbmu \up = \dmu \up + [V_{\mu} , \up ] + \{ A_{\mu} , \up \} .
\label{covdiv}
\end{eqnarray}
   The interaction with electromagnetic field $\cal{A}_{\mu}$ can be
introduced into chiral Lagrangians by using the substitution
$$
     V_{\mu} \rightarrow V_{\mu}  + ie Q \cal{A}_{\mu} .
$$
   The terms containing the matrices $\M = m(U^+ m_0 + m_0 U)$ and
$\Mb = m( m_0 U^+ + U m_0)$, $m$ being the average constituent quark mass
and $m_0 = \diag(m^0_u, m^0_d, m^0_s)$ is the mass matrix of current
quarks, take into account the additional contributions from the quark mass
expansion;
$M = diag( \chi^2_u, \, \chi^2 _d, \, \chi^2_s ) , \;
\chi^2_i = -2m^0_i F^{-2}_0 \qq $,
where $\qq$ is the quark condensate.
   Here we restrict ourselves to terms up to order $m_0$.
   The Goldberger-Treiman current quark mass splitting can be taken
into account in the bosonized Lagrangians (\ref{4a})-(\ref{4c})
using substitution $U \rightarrow \widetilde U = U + m_0/m$.

      We introduced also in (\ref{4c}) the factors $\Z_{9,10}$ to distinguish
explicitly two terms which correspond to the terms with the structural
constants $L_{9,10}$ of the Gasser-Leutwyler \cite{gas-leutw} representative
form for effective chiral Lagrangian with fourth-order derivative terms.
      For convenience, the factors $\Z_{9,10}$ are defined in a such way
that
 $$
	\Zz = \Zzz = \frac{48 \pi^2 }{N_c} L_{9}
		   = -\frac{96 \pi^2 }{N_c} L_{10} = 1,
 $$
 if $L_{9} = -2L_{10} = N_c / (48 \pi^2 )$ are the standard values
 of $L_{9,10}$ obtained from the direct calculation of the quark determinant.

    The corresponding meson currents have the form in the pseudoscalar
sector \cite{91-08}:
 \begin{eqnarray}
    J_{L\,\mu}^{a\,(div)} &=&
    i\, \frac{F^2_0}{8} \tr ( \lambda^a \dmu U \, \widetilde U^+) + \hc,
 \label{3a}
 \\
    {J}^{a\,(div)}_{L} &=&
    \frac{F^2_0}{4} \, m \, R \, \tr (\, \lambda^a \up)
  + \frac{F^2_0}{8m} \, \tr \left[ \lambda^a (\dii \up + 2 \, \up \M) \right],
 \label{3b}
 \\
    {J}^{a\,(fin)}_{L\, \mu} &=&
    {i \, N_c \over 64\pi^2} \, \tr \bigg\{ \lambda^a
    \bigg[ \frac13 \, \dnu U \, \dmu \up \dnuup U \widetilde U^+
  - (\M \dmu U + \dmu U \Mb )\, \up + \frac1{6m^2} \Big( \big(\M \dmu \dii U
 \nonumber
 \\
 &&+ \dmu \dii U \Mb + \dnu \M \dmu \dnuup U + \dmu \dnuup U \, \dnu \Mb
   + \dii \M \, \dmu U + \dmu U \, \dii \Mb  \big) U^+
 \nonumber
 \\
 &&         + (\M \dnuup U  + \dnuup U \Mb ) \dmu\dnu U^+
	    + \dnu U \dnuup U^ + \dmu \M \Big) \bigg] \bigg\} + \hc ,
 \label{3c}
 \\
 {J}^{a\,(fin)}_{L} &=&
 -{N_c \over 192\pi^2m} \tr \bigg\{ \lambda^a
  \bigg[-\dmu (\dnu U^+ \dmuup U \dnuup U^+) + U^+ \dii\M + \dii\Mb U^+
 \nonumber
 \\
 &&     -3\Big( m^2 (\up \dmu U \dmuup \up + \dmu\up  \dmuup U  \up)
	      - \dii \up \M - \Mb \dii \up
 \nonumber
 \\
 &&     - \dmu U^+ \dmuup \M - \dmu U^+ \Mb \dmuup \Big)
	+ \frac12 \up \Big(2 \{\M,\dmu U \dmuup \up \} + 2 \dmu U \M \dmuup U^+
 \nonumber
 \\
 &&     + \dmu \dnu U \dmuup \dnuup U^+ + \dmuup U \dii \dmu U^+
	+ \dii \dmu U \dmuup U^+ \Big)
 \nonumber
 \\
 &&+ \frac12 \Big(
   2 \{\dmu U^+ \dmuup U,\Mb \} + 2\dmu U^+ \Mb \dmuup U
      + \dmuup \dnuup U^+ \dmu \dnu U
 \nonumber
 \\
 &&+ \dii \dmu U^+ \dmu U
      + \dmuup U^+ \dii \dmu U \Big) U^+
	  \bigg]
	  \bigg\} ,
 \label{3d}
 \\
 J^{a(fin,\gamma)}_{L\, \mu} &=& - \frac{eN_c}{64 \pi^2} tr
  \bigg \{ \lambda^a\bigg[ \Zz \frac{2}{3} {\cal{F}}_{\mu \nu} \widetilde U
  [Q,\dnuup \up ] + \frac{1}{3} \dnu \bigg( {\cal{F}}_ {\mu \nu}
  (\frac{1}{m^2}Q\M - \Zzz \widetilde U Q \widetilde U^+ ) \bigg)
 \nonumber
 \\
 && - \frac{1}{6m^2} \bigg(
 U \Big( {\cal{F}}_{\mu \nu} [ Q,\dnuup \dii U^+ ]
 + \partial_{\alpha} {\cal{F}}_{\mu \nu} [Q,\partial^\alpha \dnuup U^+ ]
 + \dii {\cal{F}}_{\mu \nu} [Q,\dnuup U^+] \Big)
 \nonumber
 \\
 && + 2 \partial_{\alpha}{\cal{F}}_{\mu \nu} ( \partial^{\alpha} U [Q,\dnuup
U^+]
 + Q \partial^{\alpha} U \dnuup U^+ )
 - 2 \partial^{\alpha} {\cal{F}} _{\alpha \nu} ( \dmu U [ Q,\dnuup U^+]
 + Q \dmu U \dnuup U^+)
 \nonumber
 \\
 && + \dmu {\cal{F}}_{\nu \alpha} \partial^{\alpha}U [Q,\dnuup \up]
 - {\cal{F}}_{\nu \alpha} (\dmu \partial^{\alpha} U [Q,\dnuup \up]
 + 2 Q \dmu \partial^{\alpha} U \dnuup \up )
 \nonumber
 \\
 && + \frac{1}{12} {\cal{F}}_{ \mu \nu}
 ( 5 \dnuup \partial_{\alpha} U \partial^{\alpha} \up Q
 + 12 \dnuup \partial_{\alpha} U Q \partial^{\alpha} \up
 - 41 \partial^{\alpha} U \dnuup \partial_{\alpha} \up Q )
 \nonumber
 \\
 && - \frac{1}{12} \dnuup {\cal{F}}_{\mu \nu} (19 \partial_{\alpha}
 U \partial^{\alpha} \up Q + 5 \partial_{\alpha} U Q
 \partial^{\alpha}\up ) + \frac{1}{6} \partial^{\alpha}
 {\cal{F}}_{\alpha \nu} U [ Q,\dmu \dnuup \up ]
 \nonumber
 \\
 && - 2 {\cal{F}}_{\mu \nu} ( Q \dnuup U \dii \up
    - \{ Q,\dii U \} \dnuup \up )
 \nonumber
 \\
 && - 2 {\cal{F}}_{\mu \nu} U \Big( \dnuup \up \{ \Mb,Q \}
    - \{ \M,Q \} \dnuup \up  + \M \dnuup \up Q - Q \dnuup \up \Mb \Big)
    \bigg) \bigg] \bigg \}
 %
 + \hc ,
 \label{3e}
 \\ \vspace*{0.5cm}
 J^{a(fin,\gamma)}_L &=& \frac{ieN_c}{64 \pi^2 m} tr \bigg \{ \lambda^a
 \bigg[ \Zz \frac{2}{3} \dmuup \bigg({\cal{F}}^{\gamma}_{\mu \nu}[Q,\dnuup \up]
\bigg)
 \nonumber
 \\
 && + \frac{1}{6m^2} \bigg( \dmuup \Big( 2 {\cal{F}}_{\mu \nu}
      \big( \{Q,\Mb \} \dnuup \up -  \dnuup \up \{Q,\Mb \}
    + Q \dnuup \up \M - \Mb \dnuup \up Q \big)
 \nonumber
 \\
 && + \dii {\cal{F}}_{\mu \nu} [Q, \dnuup \up ]
 + \frac{7}{6} \partial^\alpha  {\cal{F}}_{\alpha \nu} [ Q,\dmu \dnuup \up ]
 \Big )
 + \frac{1}{6} \dmuup {\cal{F}}_{\mu \nu} \big( \up [ Q,\dnuup \M ]
 + [ Q,\dnuup \Mb ] \up \big)  \bigg )
 \nonumber
 \\
 && + \frac{1}{3} {\cal{F}}_{\mu \nu} \bigg ( \up ( \{ Q,\dmuup \up \dnuup U \}
 - \dmuup \up Q \dnuup U )
 \nonumber
 \\
 && + ( \{ Q, \dmuup U \dnuup \up \} - \dmuup U Q \dnuup \up ) \up \bigg )
\bigg ] \bigg \}
 %
 .
 \label{3f}
 \end{eqnarray}
   Here we write down only the terms relevant for the description of
the decays under consideration in the present paper.

   Meson currents,
$J^{a\,(fin,\gamma)}_{L_\mu}$ and $J^{a\,(fin,\gamma)}_L$,
containing the electromagnetic field strength tensor
$\cal{F}_{\mu\nu}=\partial_\mu \cal A_\nu - \partial_\nu \cal A_\mu$ ,
describe the electromagnetic-weak transitions with the emission of
``structural'' photons.
   These currents originate from the so called ``nonminimal'' part
$\L^{(fin)}_2$ (\ref{4c}) which vanish when vector and axial-vector
collective fields become equal to zero.
   The nonminimal Lagrangian $\L^{(fin)}_2$ contains both $p^4$ and
$p^6$ higher-order derivative terms which play an important role in
the description of the radiative kaon decays discussed in this paper.
   The emission of the usual bremsstrahlung photons can be included
when using the standard replacement
$\dmu * \rightarrow \dmu *  + ie {\cal A}_{\mu} [ Q, * ]$.
   The contributions of the gluon and electromagnetic penguin
operators are determined by the parameter
 $$
 R = {{\qq} \over {mF_0^2}} .
 $$

    The parameters $\chi_i^2$, $m_i^0$ and $m$ can be fixed by the spectrum
of pseudoscalar and vector mesons.
    Here we use the value $m=380 \, \mev$ and the relation
 $m_s^0 = \widehat{m}_0 \,\, \chi_s^2 / m_\pi^2$, where
 $\widehat{m}_0 \equiv (m_u^0+m_d^0)/2 \approx 5 \, \mev$,
 $\chi_u^2 = 0.0114 \,\gev^2,\; \chi_d^2 = 0.025\,\gev^2,$ and
 $\chi_s^2 = 0.47\,\gev^2. $
    Taking into account the additional Goldberger-Treiman
 contribution to $F_{K,\pi}$ arising from current quark mass
 splitting
 $$
 F_{K,\pi}=F_0\bigg[1 + { {\overline{m}^0_u +\overline{m}^0_{d,s}} \over {2} }
 \bigg(1 - { {2mF_0^2} \over {\qq} }\bigg ) \bigg] ,
 \qquad
 \overline{m}_i^0 = { {m_i^0}\over {m} } ,
 $$
 the value $F_0=89.8\,\mev$ can be obtained.

    Similarly, by applying the bosonization procedure \cite{91-08}
 to the anomalous Wess-Zumino part of the effective meson Lagrangians
 \cite{WZ}, which is related to the phase of the quark determinant, one
 obtains the Wess-Zumino electromagnetic-weak current
 \begin{eqnarray}
 %
 %
     J^{(WZ,\gamma)a}_{L \alpha} &=&
	\frac{i N_c}{48 \pi^2} \, \varepsilon_{\alpha\beta \mu \nu}\,
	 tr \, \Big \{ \lambda^a \Big [e {\cal{A}}^{\beta}
	    \Big( ( U Q \up L^\mu + L^\mu Q ) L^{\nu}
 \nonumber
 \\
 && + U R^\mu R^\nu Q \up - \{ Q,\dmuup L^\nu \} - \dmuup (U Q \up L^\nu )
\Big)
 \nonumber
 \\
 && + e \dnuup{\cal{A}}^\beta (2 \{Q,L^\mu \} + U R^\mu Q \up + UQ \up L^\mu)
 %
 \Big] \Big\}
 \label{5b}
 \end{eqnarray}
     where $L_\mu = \dmu U  \up $ and  $ R_\mu = \up \dmu U$.

%
%
\section*{2. Form factors of semileptonic and radiative semileptonic decays}
%

     The currents (\ref{3a},\ref{3c}) describe, in particular, the decays
 $K,\pi \rightarrow \pi l \nu$.
     The hadronic part of the corresponding matrix elements usually is
 parameterized in the form
 \begin{eqnarray*}
    T_{\mu}\big( K,\pi \rightarrow \pi l \nu \big) =
				       f_+(t)(k+p)_{\mu} + f_-(t)(k-p)_{\mu},
 \end{eqnarray*}
    where k and p are the 4-momenta of the decaying and final mesons.
    The form factors $f_\pm$ depend on the invariant variable $t = (k-p)^2$
 and can be written including terms up to order of $m^0_q$ as
 \begin{eqnarray}
  f^{(\pi^+ \rightarrow \pi^0)}_+ &=&
       \sqrt{2} \big( 1 - \overline{m}^0_u - \overline{m}^0_d \big)
       \bigg[ 1 + \frac{\overline{m}^0_u + \overline{m}^0_d}{2}
     - \frac{N_cm^2}{4 \pi^2 F^2_0} \big( \overline{m}^0_u + \overline{m}^0_d
\big)
       \bigg( 1 + \frac{k^2 + p^2}{8m^2} \bigg)
 \nonumber
 \\
 &&  + \frac{3 N_c}{16\pi^2 F^2_0} \big(\overline{m}^0_u+\overline{m}^0_d \big)
\, t
       \bigg ],
 \nonumber
 \\
  f^{(\pi^+ \rightarrow \pi^0)}_- &=&
     - \sqrt{2} \frac{N_c}{48 \pi^2F^2_0}
       \big( 1 - \overline{m}^0_u - \overline{m}^0_d \big)
       \big( \overline{m}^0_u + \overline{m}^0_d \big) \big(k^2 - p^2 \big) ;
 \nonumber
 \\
  f^{(K^0 \rightarrow \pi^\pm)}_+ &=&
       \bigg(1 - \frac{\overline{m}^0_u + 2\overline{m}^0_d +
\overline{m}^0_s}{2} \bigg)
       \bigg \{1 + \frac{\overline{m}^0_u + 2\overline{m}^0_d +
\overline{m}^0_s}{4}
 \nonumber
 \\
 &&  - \frac{N_c m^2}{4 \pi^2 F^2_0} \bigg[ 2 \overline{m}^0_d
     + \frac{1}{48 m^2}
       \bigg( k^2 \big( 11 \overline{m}^0_s - 3\overline{m}^0_u + 4
\overline{m}^0_d \big)
     - p^2 \big( \overline{m}^0_s - 9 \overline{m}^0_u - 4 \overline{m}^0_d
\big) \bigg)
       \bigg ]
 \nonumber
 \\
 &&  + \frac{N_c}{16 \pi^2 F^2_0}
       \big(3 \overline{m}^0_s + \overline{m}^0_u \big) \, t \bigg \},
 \nonumber
 \\
  f^{(K^0 \rightarrow \pi^\pm)}_- &=&
       \bigg(1 - \frac{\overline{m}^0_u + 2\overline{m}^0_d +
\overline{m}^0_s}{2} \bigg)
       \bigg \{ \frac{\overline{m}^0_s - \overline{m}^0_u}{4}
     - \frac{N_c m^2}{4 \pi^2 F^2_0} \bigg [ \overline{m}^0_s -
\overline{m}^0_u
 \nonumber
 \\
 &&  + \frac{1}{48 m^2} \bigg(
	k^2 \big( 12 \overline{m}^0_d - \overline{m}^0_d - 3 \overline{m}^0_s \big)
	p^2 \big( \overline{m}^0_s - 12 \overline{m}^0_d + 3 \overline{m}^0_u \big)
       \bigg) \bigg]
     + \frac{3 N_c}{16 \pi^2 F^2_0} \overline{m}^0_u t \bigg \} \, ;
 \nonumber
 \\
  f^{(K^\pm \rightarrow \pi^0)}_+ &=&
       \frac{1}{\sqrt{2}}
       \bigg(1 - \frac{\overline{m}^0_d + 2\overline{m}^0_u +
\overline{m}^0_s}{2} \bigg)
       \bigg \{ 1 + \frac{\overline{m}^0_s + 7 \overline{m}^0_u}{8}
     - \frac{N_c m^2}{4 \pi^2 F^2_0} \bigg[ 2 \overline{m}^0_u
 \nonumber
 \\
 &&  + \frac{1}{48 m^2} \bigg( 3 k^2 \big( 3\overline{m}^0_u + \overline{m}^0_s
\big)
     +  p^2 \big( 13 \overline{m}^0_u - \overline{m}^0_s \big) \bigg) \bigg]
     + \frac{N_c}{16 \pi^2 F^2_0} \big( 3 \overline{m}^0_s + \overline{m}^0_u
\big) t
       \bigg \} \, ,
 \nonumber
 \\
  f^{(K^\pm \rightarrow \pi^0)}_- &=&
       \frac{1}{\sqrt{2}}
       \bigg(1 - \frac{\overline{m}^0_d + 2\overline{m}^0_u +
\overline{m}^0_s}{2} \bigg)
       \bigg \{\frac{\overline{m}^0_s - \overline{m}^0_u}{8}
     - \frac{N_c m^2}{4 \pi^2 F^2_0} \bigg[ \overline{m}^0_s - \overline{m}^0_u
 \nonumber
 \\
 &&  + \frac{1}{48 m^2} \bigg( k^2 \big( 3\overline{m}^0_u + 5\overline{m}^0_s
\big)
     + p^2 \big( \overline{m}^0_s - 9 \overline{m}^0_u \big) \bigg]
     + \frac{3N_c}{16 \pi^2 F^2_0} \overline{m}^0_u t \bigg \} \, .
 \label{fPl3}
 \end{eqnarray}

     For the standard parameterizations of $K^{\pm}_{l3}$ form factors
 $$
   f_{\pm}(t) =
   f_{\pm}(0) \bigg(1 + \lambda_{\pm}\frac{t}{m^2_\pi} \bigg) \, ,
 \quad
   f_0(t) \equiv f_+(t) + \frac{t}{m^2_K - m^2_\pi} \, f_-(t) =
   f_0(0) \bigg(1 + \lambda_0 \frac{t}{m^2_\pi} \bigg)
 $$
    one obtains
 $$
   \sqrt{2} \, f^{(K^\pm)}_+(0) = 1.086, \quad
   \lambda^{(K^\pm)}_+ = 0.0305 ; \qquad
   \sqrt{2} \, f^{(K^\pm)}_-(0) = - 0.371, \quad
   \lambda^{(K^\pm)}_- = - 0.0038 \; ;
 $$
 $$
   \xi = f^{(K^\pm)}_-(0)/f^{(K^\pm)}_+(0) = -0.34 ,\quad
   \lambda^{(K^\pm)}_0 = 0.0027
 $$
   which are in good agreement with experiment \cite{pdg}:
 \begin{eqnarray*}
       \xi^{exp} = -0.35 \pm 0.15, \quad
       \lambda^{exp}_+ = 0.028 \pm 0.004 , \quad
       \lambda^{exp}_0 = 0.004 \pm 0.007 \, .
 \end{eqnarray*}

    The electromagnetic-weak currents (\ref{3e}) and (\ref{5b})
 describe the vector and axial-vector form factors of the semileptonic
 radiative meson decays $K, \pi \rightarrow l \nu \gamma$ and
 $K,\pi \rightarrow \pi l\nu \gamma$.
    The form factors of the decay $K, \pi \rightarrow l \nu \gamma$
are defined  by the corresponding parameterization of the
 amplitude\\
 \begin{eqnarray*}
 T_\mu \big(K,\pi \rightarrow l \nu \gamma \big ) = \sqrt{2} e \Big [
 F_V \varepsilon_{\mu \nu \alpha \beta} k^{\nu} q^{\alpha}
 \varepsilon^\beta + i F_A \Big( \varepsilon_\mu \big(k  q \big) -
 q_\mu (k  \varepsilon ) \Big ) \Big ]
 \end{eqnarray*}
 \noindent
 where $k$ is the 4-momentum of the decaying meson, $q$ and $
 \varepsilon$ are the 4-momentum and polarization 4-vector of the
 photon.

     The form factors $F_{V,A} $ can be written including terms up to order
 of $ m^0_q$ as
 \begin{eqnarray}
 F_V^{(K,\pi)} &=& \frac{1}{8 \pi^2 F_0} ,
 \nonumber
 \\
 F^{(K,\pi)}_A &=& \frac{1}{4 \pi^2 F_0} \bigg[ \Zz (1-\overline{m}_u
-\overline{m}_{s,d})
 - {1 \over 2}
   \Zzz \bigg( 1 - { {2\overline{m}_u +\overline{m}_{s,d}} \over {3} }\bigg)
					 \bigg ].
 \label{FVA}
 \end{eqnarray}
     The theoretical value of the ratio $\gamma \equiv F_A / F_V = 1$ arising
 from (\ref{FVA}) in the chiral symmetry limit ($m_0 =0$) is seemingly in
 disagreement with the experimental results on the ratio $\gamma$ from
 $\pi \rightarrow e \nu \gamma$ decay:
 \begin{eqnarray*}
	  \gamma = \left\{ \begin{array}{ll}
		   0.25 \pm 0.12 \quad \cite{PIENUG-LAMPF}, \\
		   0.41 \pm 0.23 \quad \cite{PIENUG-bolotov}. \\
 \end{array} \right.
 \end{eqnarray*}
     Clearly, $m_0$-corrections are too small to improve the description
 of the experiments.

     This problem can be solved if one takes into account $\pi A_1$-mixing,
 arising from covariant divergences (\ref{covdiv}) in the kinetic part of the
 effective Lagrangian (\ref{4a}).
     The role of $\pi A_1$-mixing was already discussed in
 ref.\cite{SJPN-volkov} where the linear realization of chiral symmetry was
 considered.
     The results of ref.\cite{SJPN-volkov} can be easily reproduced in
 nonlinear parameterization.
     Indeed the required field and decay constant redefinitions
 \begin{eqnarray*}
 \Phi \rightarrow Z^{-1}_{\pi A_1} \widetilde {\Phi} , \qquad
 F_0 \rightarrow Z^{-1}_{\pi A_1} \widetilde{F}_0 , \qquad
 V_{\mu} \rightarrow \frac{g^0_V}{\sqrt{1+ \tilde{\gamma}}}
\widetilde{V}_{\mu},
 \qquad
 A_{\mu} \rightarrow \frac{g^0_V}{\sqrt{1- \tilde{\gamma}}} \widetilde{A}_{\mu}
    + \frac{i}{\sqrt{2} \widetilde{F}_0} \big(1 - Z^2_{\pi A_1} \big)
			\partial_{\mu} \widetilde {\Phi},
 \end{eqnarray*}
necessary for removing $\pi A_1$-mixing, lead in the pure pseudoscalar meson
sector to a replacement of former covariant derivatives by the expressions
   \footnote{In general, this replacement breaks the nonlinear
transformation properties of the covariant derivatives for
pseudoscalar meson matrix $U$, so the redefinitions (\ref{redef}) can
be treated only as some formal procedure for the simplification the
meson amplitudes calculations with taking into account the
intermediate $\pi A_1$-mixing vertex.
   The bosonized nonlinear effective Lagrangian of strong interaction
and weak and electromagnetic-weak currents for pseudoscalar sector, obtained
from quark determinant after reducing vector and axial-vector degrees
of freedom, will be considered in detail elsewhere in the special
publication.}
\begin{eqnarray}
\Dmu U = \dmu U - \frac{i}{\sqrt{2} \widetilde{F}_0}
                  \big(1 - Z^2_{\pi A_1} \big)
                  \{ \partial_{\mu} \widetilde {\Phi} ,U \} , \qquad
\Dbmu \up = \dmu \up + \frac{i}{\sqrt{2} \widetilde{F}_0}
                       \big(1 - Z^2_{\pi A_1} \big)
                       \{ \partial_{\mu} \widetilde {\Phi} ,U^+ \} .
\label{redef}
\end{eqnarray}
    Here $m^0_V$ and $g^0_V$ are the bare mass and coupling constant of vector
gauge field; the factor $Z_{\pi A_1}$ is determined by the relations
$$
Z^2_{\pi A_1} \equiv 1 - \bigg( \frac{g^0_V \widetilde{F}_0}{m^0_V} \bigg)^2
              = \frac{m^2_{\rho}}{m^2_{A_1}}
                \frac{1+ \tilde{\gamma}}{1- \tilde{\gamma}},
$$
   and $\tilde{\gamma} = \Zzz \frac{N_c (g^0_V)^2}{48\pi^2}$ is the
additional factor originating from the corresponding term of $\L^{(fin)}_2$
(\ref{4c}).
    The special choice $Z^2_{\pi A_1} =1/2$ leads to the KSFR relation
$m^2_{\rho} = 2g^2_{\rho \pi \pi}\widetilde{F}_0^2$, where
$m_{\rho} = m^0_V/\sqrt{1+ \tilde{\gamma}}$ and
$g_{\rho \pi \pi} = g^0_V/\sqrt{1+ \tilde{\gamma}}$.

    Redefinitions of type (\ref{redef}) lead to the appearance the general
factor $Z^2_{\pi A_1}$ in the expression for the form factor $F_A$ and
the ratio $\gamma$ agrees with the experimental data on
$\pi \rightarrow e \nu \gamma$ decay.
    In this way it proves to be possible to remove also the inselfconsistency
in the description of ratio $\gamma$ and pion polarizability which arises
seemingly in the pseudoscalar sector of effective chiral Lagrangian with the
terms corresponding to the structural constants $L_{9,10}$,
respectively $\Z_{9,10}$, of Gasser-Leutwyler representation
\cite{gas-leutw} (see the detailed discussion of this problem, for
example, in ref. \cite{L9L10-donoghue,electroweak-holstein,semilept-bijnens}).

    The amplitude of $K, \pi \rightarrow \pi l \nu \gamma$ decay is
parameterized as
\begin{eqnarray*}
T^\mu ( K,\pi \rightarrow \pi l \nu \gamma)
 &=& \sqrt{2} e \Big [i \varepsilon^{\mu \nu \alpha \beta}
\varepsilon_\nu \Big (h_A k_\alpha q_\beta + h_A^{'}p_{\alpha} q_{\beta} \Big)
+ h_V \Big ( \varepsilon_\mu (kq) - q_\mu (k \varepsilon ) \Big )
\\ \nonumber
&& + \Big( (p \varepsilon)(kq) - (k \varepsilon )(pq) \Big)
\Big( h^{'}_{V} p_{\mu} + h^{''}_{V} k_{\mu} + h^{'''}_{V} q_{\mu} \Big) \Big],
\end{eqnarray*}
where $p$ is 4-momentum of $ \pi^0$ in the final state.
    The corresponding form factors can be written in the same approximation as
\begin{eqnarray*}
 h^{(\pi^+)}_{A} &=& h^{'(\pi^+)}_{A} = \frac{1}{8 \pi^2 F^2_0} ,
\\
 h^{(\pi^+)}_V &=& \frac{1}{8 \pi^2 F^2_0} \bigg \{ \Zz \bigg(1- \frac{5
\overline{m}^0_u + 4 \overline{m}^0_d}{3} \bigg) + \frac{1}{24m^2} \Big[
10\big(k^2 + 2
p^2\big) - 23 (kp)
- 6(kp) + 12 (pq) \Big] \bigg \},
\\
h^{'(\pi^+)}_{V} &=& - 2 h^{''(\pi^+)}_{V} = - \frac{1}{16 \pi^2 F^2_0 m^2},
\qquad
h^{'''(\pi^+)}_{V} = 0 ;
\\
h^{(K^+)}_A  &=& - \frac{1}{3} h^{'(K^+)}_A = \frac{1}{16 \pi^2
F^2_0},
\\
h^{(K^+)}_V &=& \frac{1}{16 \pi^2 F^2_0} \bigg \{ \Zz \bigg(1 - \frac{13
\overline{m}^0_u + 8 \overline{m}^0_s}{6} \bigg) + \frac{1}{24 m^2} \Big[ 10
\big(k^2 +
2 p^2 \big) - 23 \big(k  p \big)
\\
&& - 3 ( 2 - \overline{m}^0_u + \overline{m}^0_s ) \big( k  q \big )
+  \Big ( 28-3 \big(\overline{m}^0_u - \overline{m}^0_s \big ) \Big )
\big( p  q \big) \Big ] \bigg \},
\\
 h^{'(K^+)}_{V} &=& - \frac{7}{2} h^{''(K^+)}_{V} = - \frac{7}{192 \pi^2 F^2_0
m^2} ,
\qquad
 h^{'''(K^+)}_{V} = - \frac{1}{192 \pi^2 F^2_0 m^2} \bigg (1-
\frac{3}{2} \big(\overline{m}^0_u - \overline{m}^0_s \big) \bigg ) ;
\\
 h^{(K^0)}_{A} &=& - h^{'(K^0)}_{A} = -\frac{1}{8 \sqrt{2} \pi^2 F^2_0},
\\
 h ^{(K^0)}_{V} &=& \frac{1}{8 \sqrt{2} \pi^2 F^2_0} \bigg \{ \Zz \bigg(1-
\frac{5\overline{m}^0_u + 3 \overline{m}^0_d + \overline{m}^0_s}{3} \bigg) +
\frac{1}{24m^2}\Big [16m^2_K + 14m^2_\pi
\\
&& - \Big (2 + 3 \big (\overline{m}^0_u - \overline{m}^0_s \big ) \Big ) \big(
p
q \big) - 6  \big(1 - \overline{m}^0_u - \overline{m}^0_s \big) \big(k  p
\big) + 3 \big( \overline{m}^0_u - \overline{m}^0_s \big) \big( k  q\big)
\Big ] \bigg \},
\\
h^{'(K^0)}_{V} &=& -\frac{4}{5} h^{''(K^0)}_{V} = - \frac{1}{24 \sqrt{2} \pi^2
F^2_0 m^2} ,
\qquad
 h^{'''(K^0)}_{V} = \frac{5}{96 \sqrt{2} \pi^2 F^2_0 m^2} \bigg (1 +
\frac{3}{10} \big( \overline{m}^0_u - \overline{m}^0_s \big ) \bigg).
\end{eqnarray*}

%
\section*{3. $K \rightarrow \pi e^+e^-$ decays}
%

   The $K^+ \rightarrow \pi^+ e^+e^-$ decay is dominated by the one-photon
exchange diagram of fig.1a.
   This decay can also arise from a two-photon intermediate state
(see diagram of fig.1b) whose contribution is suppressed
at least by factor $\alpha = 1/137$ compared with one-photon exchange
contribution.
   In the case of $K^0_L \rightarrow \pi^0 e^+e^-$ decay, the
two-photon mechanism can compete with one-photon exchange since
the latter is forbidden if $CP$ was conserved and the
transition $K^0_L \rightarrow \pi^0 \gamma^*$ is caused by $CP$
violation.
   Since $K^0_L$ consists of a $CP$-odd state $K^0_2$ with a small admixture
of a $CP$-even state $K^0_1$,
$$
    K^0_L \approx K^0_2 + \varepsilon K^0_1
\qquad
    \big( |\varepsilon| = 2.28 \cdot 10^{-3} \big),
$$
two possible $CP$-violating contributions to one-loop exchange exist.
   The first one is the decay of the $CP$-even state
$K^0_1 \rightarrow \pi^0 \gamma^* \rightarrow \pi^0 e^+e^-$ --- ``indirect''
$CP$ violation due to the mass matrix $(K^0 \!\!-\!\! \kb)$-mixing
caused by superweak $|\Delta S|=2$ interaction.
   The second is the decay of the $CP$-odd state $K^0_2$ --- ``direct''
$CP$ violation in the $\dsone{}$ weak amplitude through the $CP$ phase
in the Kobayashi-Maskawa matrix. In the case of
$K^0_L \rightarrow \pi^0 \gamma^*$ transition the amplitude of the direct
$CP$ violation can be comparable to that of the indirect $CP$ violation,
in contrast to the case of $K^0 \rightarrow 2 \pi$ decays where the direct
$CP$ violation is at most about $10^{-3}$ of the indirect $CP$ violation.
   Thus, the renewal of interest in the one-photon exchange mechanism
of $K^0_L \rightarrow \pi^0 e^+e^-$ decay \cite{KPIEE-donog} -
\cite{KPIEE-bruno} happens due to the
possibility of getting new experimental information about $CP$ violation
from this process.
   The basic condition for such experiments is that the contribution of
the $CP$-violating one-photon exchange mechanism to the amplitude of this
decay must exceed the competing background from the $CP$-allowed
two-photon intermediate state.

   It is convenient to parameterize the amplitude of the $K(k) \rightarrow
\pi(p) \, e^+(p_+) \, e^-(p_ -)$ decay in the following form:
\begin{eqnarray*}
T(K \rightarrow \pi e^+ e^-) &=& e^2\widetilde{G}f_
{K \rightarrow \pi
\gamma^*}(q^2) \Big [ q^2(p+k)_\mu - \Big( q  (p+k) \Big) q_\mu \Big ]
\frac{1}{q^2} \bar{u}(p_-) \gamma_\mu v(p_+) ,
\end{eqnarray*}
   where $q = k-p = p_+ + p_-$ is the 4-momentum of virtual
photon and $ f_{K \rightarrow \pi \gamma^*}$ is the dimensionless
form factor of the $K(k) \rightarrow \pi(p) \gamma^*(q)$ transition:
\begin{eqnarray*}
T(K \rightarrow \pi \gamma^*) = e \varepsilon_\mu \widetilde{G}
    f_{K \rightarrow \pi \gamma^*}
    \Big [ q^2(p+k)_\mu - \Big( q  (p+k) \Big) q_\mu \Big ]
\end{eqnarray*}
with $\varepsilon_\mu$ being the photon polarization 4-vector.
   The width of $K \rightarrow \pi e^+ e^-$ decay is connected with
the form factor $f_{K \rightarrow \pi \gamma^*}$ by the relation
\begin{eqnarray*}
\Gamma (K \rightarrow \pi e^+  e^-) = e^4 \widetilde{G}^2 \frac{m_K}{12
\pi^3} \int^{(m^2_K+m^2_\pi)/(2m_K)}_{m_\pi}dE_\pi(E^2
_\pi-m^2_\pi)^{3/2}|f_{K \rightarrow \pi\gamma^*}|^2,
\label{ast1}
\end{eqnarray*}
where $E_\pi = (m^2_K + m^2_\pi - q^2)/(2m_K)$ is the pion kinetic
energy in the kaon rest frame (the electron and positron masses
are neglected).

    At the tree level it is obvious that bremsstrahlung emission of
photons does not contribute to the $K^0 \rightarrow \pi^0 \gamma^*$
transition (diagrams of fig.2a-c).
    For the $K^+ \rightarrow \pi^+\gamma^*$ transition, the
cancellation of diagrams of fig.2a-c originates from the general
properties of gauge invariance and chiral symmetry.
    The bremsstrahlung emission of photons gives nonzero contributions
to the transition $K \rightarrow \pi \gamma ^*$ only at the one-loop
level \cite{KPIEE-ecker}.

    To simplify the meson loop calculations we used the method applied
in ref. \cite{KPIEE-ecker} where only those diagrams (fig.2d) which
can lead to terms in the amplitudes proportional to
$q^2(p+k)_\mu$ were considered.
    The gauge invariant amplitude of $K \rightarrow \pi \gamma^*$
transition can then be restored by subtraction from the results obtained
from each corresponding one-loop diagram their values at $q^2=0$.
    To fix $UV$-divergences, we used the results of special superpropagator
regularization ($SP$) method \cite{SPvolkov} which is particularly
well-suited for treating loops in nonlinear chiral theories.
    The result is equivalent to dimensional regularization to one loop,
the difference being that the scale parameter $\mu$ is no longer free but
fixed  by the inherent scale of the chiral theory, namely $\mu = 4 \pi F_0$,
and $UV$ divergences have to be replaced by a finite term through the
substitution.
\begin{eqnarray*}
({\cal{C}}- 1/ \varepsilon) \rightarrow  {\cal{C}}_{SP}
    = 2 {\cal{C}} + 1
      + \frac{1}{2} \bigg[ \frac{d}{dz} \Big(\ln \Gamma^{-2}(2z+2) \Big)
                    \bigg]_{z=0} = - 1+4 {\cal{C}} \approx 1.309,
\end{eqnarray*}
where  ${\cal{C}} = 0.577$ is the Euler constant and $\varepsilon =
(4-D)/2.$

    It is worth noting that both at the tree and meson-loop level the
involved diagrams are closely connected with the electromagnetic form
factors of $\pi$ and $K$ mesons which play the dominant role in
the description of the transitions $K \rightarrow \pi \gamma^*$.
    The electromagnetic squared radii are defined as coefficients of
$q^2$-expansion of the electromagnetic form factor $f^{em}_M(q^2)$,
$M = \pi , K$:
\begin{eqnarray*}
    <M(p_2)|V^{em}_{\mu}|M(p_1)> = f^{em}_M(q^2)(p_1 - p_2)_{\mu},
\\
    f^{em}_M(q^2) = 1 + \frac{1}{6} <r^2_{em}>_M q^2 + \cdots .
\end{eqnarray*}
    Being restricted only by pion loops one gets in the
$SP$-regularization for the electromagnetic squared radii of $\pi$ and
$K$  mesons \cite{EMRADvolkov,BOOKvolkov}
\begin{eqnarray*}
<\!\!r^2_{em}\!\!>^{(loop)}_{\pi^+}
      &\approx& <\!\!r^2_{em}\!\!>^{(loop)}_{K^+}
       \approx -<\!\!r^2_{em}\!\!>^{(loop)}_{K^0}
\\
      &\approx & - \frac{1}{(4 \pi F_0)^2} \Big[ 3 {\cal{C}}
                 + \ln \Big( \frac{m_ \pi}{2 \pi F_0} \Big)^2 - 1 \Big]
                 = 0.063 \fm^2 \, .
\end{eqnarray*}
     Because the main contribution to this value arises from
logarithm term, the kaon loop contributions, containing the small
logarithm $\ln \Big(m_K / (2 \pi F_0) \Big) ^2$, can be neglected.
     At the Born level the corresponding contributions to the
electromagnetic squared radii of $\pi$ and $K$ mesons originate from the
nonminimal $p^4$-part of the effective Lagrangian (\ref{4c}):
\begin{eqnarray*}
<\!\!r^2_{em}\!\!>^{(Born)}_{\pi^+} &=& \frac{N_c}{4 \pi^2 F^2_0} \Zz
(1 - \overline{m}^0_u - \overline{m}^0_d)
\bigg( 1 - \frac{2\overline{m}^0_u + \overline{m}^0_d}{3} \bigg) = 0.352 \fm^2,
\\
<\!\!r^2_{em}\!\!>^{(Born)}_{K^+} &=& \frac{N_c}{4 \pi^2 F^2_0} \Zz
(1 - \overline{m}^0_u - \overline{m}^0_s)
\bigg(1 - \frac{2\overline{m}^0_u + \overline{m}^0_s}{3} \bigg) = 0.218 \fm^2,
\\
<\!\!r^2_{em}\!\!>^{(Born)}_{K^0} &=& - \frac{N_c}{12 \pi^2 F^2_0} \Zz
(1 - \overline{m}^0_d - \overline{m}^0_s) ( \overline{m}^0_s -
\overline{m}^0_d) = -0.025 \fm^2.
\end{eqnarray*}

    The total values of electromagnetic squared radii corresponding to the
sum of loop and Born contributions
$$
<\!\!r^2_{em}\!\!>_{\pi^+} = 0.413 \fm^2, \qquad
<\!\!r^2_{em}\!\!>_{K^+} = 0.281 \fm^2, \qquad
<\!\!r^2_{em}\!\!>_{K^0} = -0.088 \fm^2
$$
are in a good agreement with the experimental data
$$
<\!\!r^2_{em}\!\!>^{(exp)}_{\pi^+} = (0.439 \pm0.030 ) \fm^2 \, \cite{dally},
$$
$$
<\!\!r^2_{em}\!\!>^{(exp)}_{K^+} = ( 0.28 \pm0.05) \fm^2 \, \cite{dally},
\qquad
<\!\!r^2_{em}\!\!>^{(exp)}_{K^0} = ( -0.054 \pm 0.026 ) \fm^2 \, \cite{molzon}.
$$

    The one-loop contributions to amplitudes of $K \rightarrow \pi
\gamma^*$ transitions are given by
\begin{eqnarray*}
f^{(loop)}_{K^+\rightarrow \pi^+ \gamma^*} &=& - \Big[ (-\xi_1 + \xi_2
+ \xi_3) + \xi_4 + 4(\xi_5 + 4\xi_8) R \Big] ({\cal{F}}_K +
{\cal{F}}_\pi),\\
f^{(loop)}_{K^0\rightarrow \pi^0 \gamma^*} &=& \frac{1}{m^2_K -
m^2_\pi} \frac{1}{\sqrt{2}} \bigg \{ \Big[ \Big( 2(-\xi_1 + \xi_2 + \xi_3)
- \xi_4 + 6\xi_7 + 8( \xi_5 + \xi_8)R \Big) m^2_K\\
&& - 2\Big( (-\xi_1 + \xi_2 + \xi_3) + 6\xi_7 - 2\xi_4 +
4(\xi_5 + 2\xi_8) R \Big) m^2_\pi - 24 \mu^2 \xi_8 R^2 \Big]
\cal{F}_K\\
&& - \Big[ 3 ( \xi_4 - 2\xi_7 + 8 \xi_8 R ) m^2_\pi -
24 \mu^2 \xi_8 R^2 \Big] \cal{F}_\pi \bigg \},
\end{eqnarray*}
 where
\begin{eqnarray*}
{\cal{F}}_{K,\pi} = \frac{1}{192 \pi^2} \bigg \{ -\frac{1}{2}
\bigg({\cal{C}}_{SP} + \ln \frac{m^2_{K,\pi}}
{16 {\pi}^2 F^2_0} - 1 \bigg) + \frac{5}{6}
+ \bigg( \frac{4 m^2_{K,\pi}}{q^2} - 1 \bigg) \, J \bigg(
\frac {4 m^2 _{K,\pi}}{q^2} \bigg) - \frac{4 m^2 _{K,\pi}}{q^2} \bigg \}
\end{eqnarray*}
and
\begin{eqnarray*}
J( \zeta) = \left\{ \begin{array}{lll}
            \frac{y}{2} \ln \Big(\frac{y+1}{y-1} \Big),
                          & \zeta<0 , & y = \sqrt{1-\zeta};\\
 \frac{y}{2} \Big[ -i\pi + \ln \Big(\frac{1+y}{1-y} \Big) \Big],
                          & 0< \zeta <1 ;\\
 \bar{y} \arctan \bar{y}^{-1},
                          &  \zeta > 1, & \bar{y} = \sqrt{\zeta -1}.
\end{array} \right.
\end{eqnarray*}

    Unlike the bremsstrahlung emission, the structural photons give
nonzero contribution in the tree approximation of diagrams of fig.2a-c:
\begin{eqnarray*}
f^{(Born)}_{K^+\rightarrow \pi^+ \gamma^*} &=& \frac{1}{32 \pi^2} \Bigg \{
(-\xi_1 + \xi_2 + \xi_3 + \xi_4 ) \bigg \{
\Zz \bigg( 1-\frac{\overline{m}^0_s}{3} \bigg)
+ (2+\overline{m}^0_s) \frac{5 q^2 + 7 (m^2_K + m^2_\pi)}{48 m^2}
\\
&& - \frac{\overline{m}^0_s}{2(m^2_K - m^2_\pi)} \bigg[
 \Zz \bigg( m^2_K + \frac{5}{3} m^2_\pi \bigg)
- \frac {(m^2_K + 6 m^2) (m^2_K + m^2_\pi)}{8 \pi^2F^2_0} \bigg]
\bigg \}
\\
&& - 4(\xi_5 + 4 \xi_8) \bigg\{ - R \bigg( 1 +
\frac{3m^2 \overline{m}^0_s}{8 \pi^2 F^2_0} \bigg) \frac{5q^2}{48m^2} -
\frac{3(m^2_K + m^2_\pi) + 2 \overline{m}^0_s m^2_{\pi}}{6m^2}
\\
&& + \frac{\overline{m}^0_s m^2_{\pi}}{m^2_K - m^2_\pi} \bigg[ \frac{4}{3}
\bigg( R -\frac{m^2_{\pi}}{4m^2} \bigg)
+ \frac{m^2_{\pi}}{64\pi^2 F^2_0} \bigg( \frac{m^2_K + m^2_\pi}{m^2} +
12 \frac{m^4_K + m^4_\pi - m^2_K m^2_\pi}{m^4_{\pi}} \bigg) \bigg] \bigg \}
\\
&& - 4 \xi_5 \bigg\{ \frac{10}{3} R \bigg[ \Zz \bigg( 1 +
\frac{49}{80} \overline{m}^0_s \bigg) - \frac{9}{40} \bigg( 1 - \frac{7}{6}
\overline{m}^0_s \bigg) \frac{m^2} {\pi^2 F^2_0} - \frac{7}{160} \bigg( 1 +
\frac{6m^2}{7\pi^2 F^2_0} \bigg) \frac{m^2_K + m^2_\pi}{m^2} \bigg]
\\
&& - \frac{\overline{m}^0_s}{384 \pi^2 F^2_0 (m^2_K - m^2_\pi)} R \Big[80 m^2_K
m^2_\pi + 43 m^4_\pi - 11 m^4_K
+ 48 m^2(3m^2_K + 11m^2_\pi)  \Big] \bigg\}
\\
&& - 16 \xi_8 \bigg \{ \frac{4}{3} R \bigg[
\Zz \bigg( 1- \frac{1}{64} \overline{m}^0_s \bigg) + \frac{9}{32} \bigg( 1
+ \frac {\overline{m}^0_s}{3} \bigg) \frac{m^2}{\pi^2 F^2_0} - \frac{7}{64}
\bigg( 1
- \frac{3m^2}{7 \pi^2 F^2_0} \bigg) \frac{m^2_K + m^2_\pi}{m^2}
\bigg]
\\
&& - \frac{\overline{m}^0_s}{m^2_K - m^2_\pi} \bigg [ 2R^2 m^2
\bigg( 1-\frac{3}{32} \, \frac{m^2_\pi + 6 m^2}{\pi^2 F^2_0} \bigg)
\\
&& - \frac{1}{384 \pi^2 F^2_0} \, R \Big(35m^4_K - 32m^2_K
m^2_\pi - 19m^4_\pi + 24m^2 (3m^2_K - 7m^2_\pi) \Big) \bigg] \bigg \} \Bigg \},
\\
f^{(Born)}_{K^0\rightarrow \pi^0 \gamma^*} &=& \frac{1}{96 \sqrt{2}
\pi^2} \bigg \{ (- \xi_1 + \xi_2 + \xi_3 - 2 \xi_4 + 6 \xi_7)
\Zz \overline{m}^0_s \, \frac{m^2_K}{ m^2_K - m^2_\pi}
\\
&& + 4 ( \xi_5 - 2 \xi_8 ) \, \bigg[ R \, \Zz \Big[ \bigg( 1 +
\frac{3}{2} \overline{m}^0_s \bigg) + \frac{3 m^2 \overline{m}^0_s}{8 \pi^2
F^2_0}
\\
&&+ \frac {\overline{m}^0_s m^2_{\pi}}{m^2_K - m^2_\pi} \bigg( 1 -
\frac{2m^4_\pi -
m^4_K + 12m^2 m^2_\pi}{16 \pi^2 F^2_0 m^2_{\pi}} \bigg) \Big]
- \frac {\overline{m}^0_s m^2_K m^2_\pi}{4m^2(m^2_K - m^2_ \pi)}
\bigg ] \bigg \} .
\end{eqnarray*}
    For simplicity, we restricted ourselves by including only the dominating
$m^0_s$-contributions in the order of $m^0_q$.
    The contribution of $(\pi^0-\eta-\eta^{'})$-mixing to
$K^0 \rightarrow \pi^0 \gamma^*$ transition is small and can be neglected.

    We have considered one-photon exchange mechanism of
$K^+ \rightarrow \pi^+ e^+e^-$ decays owing to four-quark operators
(\ref{four-quark}).
    The other contribution to $K \rightarrow \pi e^+e^-$ decays arise from
the electromagnetic penguin diagram of fig.3 generating two additional
quark-lepton operators \cite{KPIEE-gilman,KPIEE-donog,KPIEE-flynn}
$$
\O_7^{'} = \frac{e^2}{4 \pi} \bar{d}_L \gamma_\mu s_L \; \bar{l} \gamma^\mu l,
\qquad
\O_8^{'} = \frac{e^2}{4 \pi} \bar{d}_L \gamma_\mu s_L \; \bar{l} \gamma^\mu
\gamma^5 l.
$$
    The operators $\O_{7,8}^{'}$ contribute to the amplitudes of
$K \rightarrow \pi e^+e^-$ transition at the tree level:
\begin{eqnarray*}
T(K \rightarrow \pi e^+ e^-) &=& \frac{e^2}{8 \pi} \gt \bigg[
c_7^{'}\;f^{(K \rightarrow \pi)}_{+}(q^2)\; (k+p)_{\mu} \;
                            \bar{u}(p_-) \gamma^\mu v(p_+)
\\
&& + c_8^{'}\;f^{(K \rightarrow \pi)}_{-}(q^2)\; (k-p)_{\mu} \;
                            \bar{u}(p_-) \gamma^\mu \gamma_5 v(p_+)
                                                                 \bigg],
\end{eqnarray*}
   where $c_{7,8}^{'}$ are Wilson coefficients corresponding to the operators
$\O_{7,8}^{'}$ and form factors $f^{(K \rightarrow \pi)}_{\pm}$ are defined as
\begin{eqnarray*}
 f^{(K^+ \rightarrow \pi^+)}_{\pm} =
                    f^{(K^0 \rightarrow \pi^{\pm})}_{\pm}
                    \big(m_u \leftrightarrow m_d \big),
\qquad
  f^{(K^0 \rightarrow \pi^0)}_{\pm} =
                   -f^{(K^{\pm} \rightarrow \pi^0)}_{\pm}
                    \big(m_u \leftrightarrow m_d \big)
\end{eqnarray*}
   (see eqs.(\ref{fPl3})).

    As the analysis of the Wilson coefficients of four-quark operators
in leading-log approximation of QCD has shown, the main contribution
to the absolute values of amplitudes of $K^+ \rightarrow \pi^+ e^+ e^-$
and $K^0_1 \rightarrow \pi^0 e^+ e^-$ decays give the nonpenguin operators
$\O_i (i=1,...,4)$ belonging to the dominating combination
$(-\xi_1 + \xi_2 + \xi_3)$ and to $\xi_4$, respectively, and the gluonic
penguin operator $\O_5$, respectively $\xi_5$.
    The contributions of the penguin operators $\O_{6,7,8}$ are small
$(\re c_{6,7,8} \ll \re c_5)$ and can be neglected.
    The contributions of the electromagnetic quark-lepton penguin operators
$\O^{'}_{7,8}$ to absolute values of decay amplitudes prove to be also small
since corresponding meson matrix elements are not enhanced compared
with nonpenguin four-quark operators unlike the $\O_5$ penguin operator
containing right-handed quark currents.

    Thus for the estimation of the $K^+ \rightarrow \pi^+ e^+ e^-$,
$K^0_S \rightarrow \pi^0 e^+ e^-$ and indirect $CP$-violating
$K^0_L \rightarrow \pi^0 e^+ e^-$ decays we can use the following values of
the parameters
\begin{eqnarray}
(-\xi_1 + \xi_2 + \xi_3) = 6.96 \pm 0.48 , \quad
                   \xi_4 = 0.516 \pm 0.025, \quad
                   \xi_5 = -0.183 \pm 0.022
\label{ksi12345}
\end{eqnarray}
    fixed as phenomenological parameters from the simultaneous analysis of
$K \rightarrow 2 \pi$ and $K \rightarrow 3 \pi$ experimental data in
\cite{DESY92-106}.
    The theoretical branching ratios  of $ K \rightarrow \pi e^+ e^-$
decays corresponding to the central values of the parameters estimates
given above are
$$
  B \big(K^+ \rightarrow \pi^+ e^+ e^- \big) = 2.4 \cdot 10^{-7}\,, \quad
  B \big(K^0_S \rightarrow \pi^0 e^+ e^-\big) = 1.4 \cdot 10^{-9} \,
$$
and
$$
  B \big(K^0_L \rightarrow \pi^0 e^+ e^- \big)_{indir} = 4.1 \cdot 10^{-12}.
$$
    The theoretical value of $B \big(K^+ \rightarrow \pi^+ e^+ e^- \big)$ is
in agreement with the experiment:
\begin{eqnarray*}
   B(K^+ \rightarrow \pi^+ e^+ e^-) =
    \left\{ \begin{array}{ll}
   (2.7 \pm 0.5) \cdot 10^{-7}\,\,\, \cite{bloch},\\
   (2.75 \pm 0.23 \pm 0.13) \cdot 10^{-7}\,\,\, \cite{alliegro}.\\
\end{array} \right.
\end{eqnarray*}

    The experimental upper limit for $K^0_S \rightarrow \pi^0 e^+ e^-$
decay is \cite{pdg}
\begin{eqnarray*}
   B(K^0_S \rightarrow \pi^0 e^+ e^-) < 4.5 \cdot 10^{-5}.
\end{eqnarray*}
    The relative contributions of tree and one-loop diagrams to the
amplitudes $f_{K\rightarrow \pi \gamma^*}$ are shown in Table \ref{1}.

    Our approach to fixation of $UV$ divergences in meson loops calculation
differs from the one of ref.\cite{KPIEE-ecker} where two estimates for
indirect $CP$-violating contribution to $K^0_L \rightarrow \pi^0 e^+e^-$
decay were obtained:
\begin{eqnarray*}
B(K^0_L \rightarrow \pi^0 e^+e^-)_{indir} = \left\{ \begin{array}{ll}
1.5 \cdot 10^{-12}, \\ 1.5 \cdot 10^{-11}. \\
\end{array} \right.
\end{eqnarray*}
    which correspond to the two possible values of renormalization scale
$w_s$ arising as a free parameter of regularization procedure based on
the usage of the fourth-order counterterms.
    The two values for the scale $w_s$ were then obtained from experimental
data on $K^+ \rightarrow \pi^+ e^+e^-$ decay as the solutions of the
corresponding quadratic equation.
    In a such approach the transitions $K^+ \rightarrow \pi^+ \gamma^*$
and $K^0 \rightarrow \pi^0 \gamma^*$ prove to be dominated by meson loops.

    As the analysis of the coefficients $c_i$ in leading-log approximation
of QCD shows, the main contribution to direct $CP$ violation comes from
the penguin diagrams.
    If we neglect the contribution of electromagnetic penguin operators,
the imaginary part of the coefficient $c_5$, responsible for the direct
$CP$ violation, can be related with the parameter $\varepsilon'$,
characterizing direct $CP$ violation in $K^0 \rightarrow 2 \pi$ decays,
from the phenomenological analysis of nonleptonic kaon decays as
\cite{DESY92-106}
\begin{eqnarray}
 |\im c_5 | = 0.053 ^{+0.015}_{-0.011} \; | \varepsilon' / \varepsilon |.
\label{Imc5}
\end{eqnarray}
    The corresponding branching ratio of direct $CP$-violating
$K^0_L \rightarrow \pi^0 e^+ e^-$ transition generated by operator $\O_5$ is:
\begin{eqnarray*}
B (K^0_L \rightarrow \pi^0 e^+ e^-)^{dir}_{\O_5} =
          3.2 \cdot 10^{-8}| \varepsilon' / \varepsilon |^2.
\end{eqnarray*}

    If we try to estimate the effect of electroweak four-quark penguin
operator $\O_{7,8}$ , the contribution of $\O_7$ may be neglected in
comparison to the dominant contribution of the operator $\O_8$.
    Because of the strong dependence of $\im c_8$ on the mass of the
$t$-quark for $m_t \ge  100 \gev$ on one hand, the contribution to direct
$CP$ violation from electroweak penguin operator becomes important for
large $m_t$.
    Using the dependence of the ratio $\eta_8(m_t) = \im c_8 / \im c_5$
on $m_t$, as derived in the papers \cite{buchalla,paschos}, and repeating
the phenomenological procedure of fixation $\im c_5$ described in
\cite{DESY92-106}, we found the connection between the contribution of
penguin four-quark operators $\O_{5,8}$ to
$B(K^0_L \rightarrow \pi^0 e^+ e^-)^{dir}_{\O_{5,8}}$
and $\varepsilon'$ shown in Table \ref{2}.
    In the case of electromagnetic quark-lepton penguin operators the
contribution of $\O^{'}_8$ may be neglected in comparison to the dominant
contribution of the operator $\O^{'}_7$.
    In analogous way, using the dependence of the ratio
$\eta^{'}_7 (m_t) = \im c^{'}_7 / \im c_5$ on $m_t$ derived in
\cite{KPIEE-flynn} we calculate the contribution of the operators
$\O^{'}_7$ to branching ratio of the direct $CP$-violating transition
$B(K^0_L \rightarrow \pi^0 e^+ e^-)^{dir}_{\O^{'}_{7}}$  together with
the total effect $B(K^0_L \rightarrow \pi^0 e^+ e^-)^{dir}_{tot}$ which
also shown also in Table \ref{2}.

%
\section*{4. $K \rightarrow \pi \pi \gamma$ decays}
%

    The amplitude of the decay $K\big( k \big ) \rightarrow \pi_1
\big (p_1 \big )\pi_2 \big( p_2 \big ) \gamma \big( q \big )$
contains two types of contributions
\begin{eqnarray*}
T \big( K \rightarrow \pi \pi \gamma \big ) = T^{IB} + T^{DE},
\end{eqnarray*}
    where $ T^{IB}$ is the inner bremsstrahlung $(IB)$ amplitude and
$T^{DE}$ describes the emission of structural photon (direct emission, $DE$).
    The $IB$ amplitudes are connected due to gauge invariance of the
electromagnetic interaction with the corresponding amplitudes of
$ K \rightarrow \pi \pi$ decays:
\begin{eqnarray}
    T^{IB} \big( K^+ \rightarrow \pi^+ \pi^{0} \gamma \big) &=&
    -e \, \varepsilon_{\mu} \, T \big( K^+ \rightarrow \pi^+ \pi^0 \big)
    \bigg( \frac{p_{1\mu}}{q \!\cdot\! p_1}
         - \frac{k_{\mu}}{q \!\cdot\! k} \bigg),
\nonumber
\\
    T^{IB} \big( K^0 \rightarrow \pi^+ \pi^{-} \gamma \big ) &=&
    -e \, \varepsilon_{\mu} \, T \big( K^0 \rightarrow  \pi^+ \pi^- \big)
    \bigg( \frac{p_{1 \mu}}{q \!\cdot\! p_1}
         - \frac{p_{2 \mu}}{q \!\cdot\! p_2} \bigg)
\label{IB}
\end{eqnarray}
    where $\varepsilon$ is the photon polarization and
$$
    T \big( K^+ \rightarrow \pi^+ \pi^{0}\big) =
                            \frac{\sqrt{3}}{2} A_2 e^{i \delta ^2_0},
\qquad
    T \big( K^0 \rightarrow \pi^+ \pi^{-}\big) =
                            \frac{1}{\sqrt{3}} A_0 e^{i \delta ^0_0}
                          + \frac{1}{\sqrt{6}} A_2 e^{i \delta ^2_0}.
$$
    Here $A_2$ and $A_0$ give the $K \rightarrow \pi \pi$ transition
amplitudes into states with isospins $I=2,0$; $\delta ^2_0$ and $\delta ^0_0$
are the $s$-wave phases arising from $\pi \pi$ final-state interactions.
    The lowest multipole transitions lead to $p$-wave $\pi \pi$ states and the
corresponding $DE$ amplitudes can be represented by the sum of magnetic $(M1)$
and electric $(E1)$ dipole transitions:
\begin{eqnarray*}
    T^{DE} = i \, e \, \varepsilon_{\mu}
             \Big[ h_{M1} \,\varepsilon_{\mu \nu \alpha \beta}
                             k_{\nu} p_{1 \alpha} q_{\beta} \,
              + i\,h_{E1} \Big( \big(q \!\cdot\! k \big) p_{1 \mu}
                              - \big(q \!\cdot\! p_1 \big) k_{\mu} \Big)
             \Big] exp (i\delta^1_1).
\end{eqnarray*}
    The $h_{M1}$ and $h_{E1}$ are the form factors of the magnetic
and electric dipole contributions respectively of structural photon
emission which have the corresponding properties under charge conjugation:
\begin{eqnarray}
        h_{M1}\big( K^+ \rightarrow \pi^+ \pi^{0} \gamma \big) =
       -h_{M1}\big( K^- \rightarrow \pi^- \pi^{0} \gamma \big) ,
\quad   h_{E1}\big( K^+ \rightarrow \pi^+ \pi^{0} \gamma \big) =
        h_{E1}\big( K^- \rightarrow \pi^- \pi^{0} \gamma \big) ;
\nonumber
\\
        h_{M1}\big( K^0 \rightarrow \pi^+ \pi^{-} \gamma \big) =
        h_{M1}\big( \kb \rightarrow \pi^- \pi^{+} \gamma \big) ,
\quad   h_{E1}\big( K^0 \rightarrow \pi^+ \pi^{-} \gamma \big) =
       -h_{E1}\big( \kb \rightarrow \pi^- \pi^{+} \gamma \big)
\label{M1E1}
\end{eqnarray}
    and $\delta^1_1$ is the $p$-wave $\pi \pi$ scattering phase shift.
    The amplitudes $A_{2,0}$ and form factors $h_{M1,E1}$ are the complex
quantities with the imaginary parts defined by the direct $CP$ violation.

    In the case of the $K^+ \rightarrow \pi^+ \pi^{0} \gamma $ decay the
$IB$ contribution is suppressed due to $| \Delta I | = 1/2$  rule
because $K^+ \rightarrow \pi^+ \pi^0$ decay is the pure $|\Delta I|= 3/2$
transition.
    On the other hand both $|\Delta I | = 3/2$ and $|\Delta I| = 1/2$
transitions contribute to $DE$ amplitude.
    In the $K^0_L \rightarrow \pi^+ \pi^- \gamma$ decay the $IB$ contribution
is suppressed due to the fact that transition $K^0_L \rightarrow \pi^+\pi^-$
is caused completely by $CP$ violation just as the decay
$K^0_S \rightarrow \pi^+ \pi^- \gamma$ dominated by $IB$ transition.
    The suppression of $IB$ amplitudes of
$K^+ \rightarrow \pi^+ \pi^0 \gamma$ and $K^0_L \rightarrow \pi^+ \pi^- \gamma$
decays gives the possibility to extract the $DE$ contribution from the
experimental data on these processes.
    The study of the direct emission of photon in
$K \rightarrow \pi \pi \gamma$ decays provides us with the important
experimental information which is relevant not only to testing of the
various chiral models but also to understanding of the mechanisms of
$CP$-violation (see, for example, \cite{K2pigam-our}-\cite{K2pigam-picciotto}
and references therein).

    At the Born level of the chiral theory the decays
$K \rightarrow \pi \pi \gamma$ are described by the diagrams of fig.4.
    The $E1$ $DE$ transitions of $K \rightarrow \pi \pi \gamma$ decays
originate from the contact diagram of fig.4a corresponding to the weak
interaction (\ref{weak-bosl}) with the currents (\ref{3a})-(\ref{3f}).
    The current contributing to the contact diagram of fig.4a for
$M1$ $DE$ transition is the Wess-Zumino electromagnetic-weak current of
the eq.(\ref{5b}).
    The $(\Phi^3 \gamma)$-vertices in the pole diagrams of fig.4b,c,d are
described by the part of anomalous Wess-Zumino electromagnetic interaction
$$
   \L^{WZ,em} = - \frac{e N_c}{48 \pi^2}\,\varepsilon^{\mu \nu \alpha \beta}\,
                  {\cal{A}}_{\mu} tr\,\Big[ Q \Big(L^\nu L^\alpha L^\beta
                + R^\nu R^\alpha R^\beta \Big) \Big] \, .
$$

    The magnetic and electric dipole form factors, corresponding to the
diagrams of fig.4, for the various channels of $K \rightarrow \pi \pi \gamma$
decays are
\begin{eqnarray*}
   h^{M1}_{K^+\rightarrow \pi^+ \pi^0 \gamma} &=&
     \frac{\widetilde{G}}{32 \pi^2 F_0}
     \bigg \{ 6(\xi_1 - \xi_2 - \xi_3 )
     \bigg[ 1 + \frac{1}{24 \pi^2 F^2_0}
     \Big( 3m^2_K + m^2_\pi - 2\big(  p_1\!\cdot\!p_2 + 2p_1\!\cdot\!q
                                    + p_2\!\cdot\!q \big) \Big) \bigg]
\\
&& - 8\xi_4 \bigg[ 1 - \frac{1}{32 \pi^2 F^2_0}
     \Big( m^2_K - 5m^2_\pi - 2\big(  3p_1\!\cdot\!p_2 + 2p_1\!\cdot\!q
                                    - p_2\!\cdot\!q \big) \Big) \bigg]
\\
&& + 2 (2 \xi_1 + 10 \xi_2 + \xi_3 - 3 \xi_6 - 12 \xi_7 )
     \bigg( 1 + \frac{m^2_K - m^2_\pi- 2\big( p_1\!\cdot\!(p_2+q) \big)}
                     {8 \pi^2 F^2_0}
     \bigg)
\\
&& - 4(\xi_5 + 4\xi_8) \bigg( 4R - \frac{m^2_K + m^2_\pi}{m^2}\bigg)
   - 2(\xi_5 - 2\xi_8) \frac{3\overline{m}^0_s m^2}{\pi^2 F^2_0} R
     \bigg\},
\\
   h^{E1}_{K^+ \rightarrow \pi^+ \pi^0 \gamma } &=&
     \frac{\widetilde{G}}{96 \pi^2 F_0}
     \bigg \{ 9(-\xi_1 + \xi_2 +\xi_3 + \xi_4) \bigg[ \Zz
   - \frac{4}{3} \Big( 1 + \frac{2}{3} \overline{m}^0_s \Big) \Zzz \bigg]
\\
&& - (-\xi_1 + \xi_2 + \xi_3 + 13 \xi_4 - 24 \xi_7) \Zz \overline{m}^0_s
   + 24 \xi_5 \Zz R \Big ( 1 + \frac{3}{4} \overline{m}^0_s \Big)
   - 24 \xi_8 \Zz R \overline{m}^0_s \bigg \};
\\
   h^{M1}_{K^0 \rightarrow \pi^+ \pi^- \gamma} &=&
     \frac{\widetilde{G}}{16 \sqrt{2} \pi^2 F_0} \bigg \{
     ( -\xi_1 + \xi_2 + \xi_3 )\bigg[ \big(7+4 \overline{m}^0_s \big)
   - \frac{m^2_K \big( 2 + \overline{m}^0_s \big)}{m^2_K - m^2_\pi}
\\
&& + \frac{1}{4 \pi^2 F^2_0} \Big( m^2_K - 5 m^2_\pi
                                  +8\big(3p_1\!\cdot\!p_2 +p_2\!\cdot\!q
                                    \big) \Big)\bigg]
\\
&& - \xi_4 \bigg[ \big( 7 + \overline{m}^0_s \big)
   - \frac{2m^2_K \big( 2 + \overline{m}^0_s \big)}{m^2_K - m^2_\pi}
   - \frac{1}{2 \pi^2 F^2_0} \Big( m^2_K + 7 m^2_\pi
                                  -4\big(3p_1\!\cdot\!p_2 +p_2\!\cdot\!q
                                    \big) \Big)\bigg]
\\
&&
   - \frac{1}{3} (2 \xi_1 + 10 \xi_2 + \xi_3 + 3 \xi_6) (2 + \overline{m}^0_s)
\\
&& - 6\xi_7 \bigg[ \frac{4}{3} (5 + \overline{m}^0_s )
   + \frac{m^2_K (2+\overline{m}^0_s )}{m^2_K - m^2_\pi}
   + \frac{m^2_\pi + p_1\!\cdot\!p_2}{\pi^2 F^2_0} \bigg]
\\
&&   - 2(\xi_5 - 2\xi_8) \frac{m^2}{m^2_K - m^2_\pi}
     \bigg [R^2 \overline{m}^0_s + 4R \frac{m^2_K}{m^2}
     \bigg( 1 + \frac{3 \overline{m}^0_s m^2}{8 \pi^2 F^2_0} \bigg)
   - \frac{m^4_K}{m^4} \bigg ]
\\
&&
   + \frac{4}{\sqrt{3} F_0^2}
     \bigg[ \frac{(cos \,\varphi - \sqrt{2} sin \,\varphi)
                  (T^8 cos \,\varphi - T^0 sin \,\varphi)}{m^2_\eta - m^2_K}
\\
&&
          + \frac{(sin \,\varphi + \sqrt{2} cos \,\varphi)
                  (T^8 sin \,\varphi + T^0 cos \,\varphi)}
                 {m^2_{\eta^{'}} - m^2_K} \bigg] \bigg \},
\\
  h^{E1}_{K^0 \rightarrow \pi^+ \pi^- \gamma} &=&
     \frac{\widetilde{G}}{48\sqrt{2} \pi^2 F_0} \bigg\{
     9 (\xi_1-\xi_2-\xi_3-\xi_4 )
     \bigg[ \bigg( 1-\frac{\overline{m}^0_s}{9} \bigg) \Zz
   - \frac{4}{3} \Zzz \bigg]
\\
&& + 4 (\xi_1 - \xi_2 - \xi_3 + 2 \xi_4 + 6 \xi_7 ) \overline{m}^0_s \Zzz
   - 24 \Big(\xi_5 + \xi_8 \overline{m}^0_s \Big) \Zz R \bigg\} \, .
\end{eqnarray*}
    Here we restricted ourselves again for simplicity to the
dominating $m^0_s$ contributions of order $m^0_q$; $\varphi = -19^o$ is the
$(\eta, \eta^{'})$-mixing angle, defined by
$$
 \eta_8 =   \eta \,\,cos\,\varphi + \eta^{'} \,\,sin\,\varphi,
\qquad
 \eta_0 = - \eta \,\,sin\,\varphi + \eta^{'} \,\,cos\,\varphi,
$$
    and $T^{8,0}$ are the amplitudes of $K^0 \rightarrow \eta^{8,0}$
transitions on the kaon mass shell:
\begin{eqnarray*}
   T^8 &=& \frac{\widetilde{G} F^2_0 m^2_K}{12 \sqrt{3}}
           \bigg\{ \big( \xi_1 - \xi_2 -(1-10/\sqrt{3})\xi_3 - 6\xi_7\big)
                   \big(6+ 7 \overline{m}^0_s \big)
         - 2\big( 2\xi_1 - 15\xi_2 - 3\xi_6 \big) \overline{m}^0_s
\\
&&       + 6\big( \xi_8 - 2\xi_5 \big) \bigg[
           6R^2 \overline{m}^0_s \frac{m^2}{m_K^2}
         - R \bigg( 8 + 15\frac{\overline{m}^0_s m^2}{\pi^2 F^2_0} \bigg)
         + \frac{2m^2_K}{m^2} \bigg ] \bigg \},
\\
   T^0 &=& \sqrt{\frac{2}{3}} \frac{\widetilde{G} F^2_0 m^2_K}{6}
           \bigg\{- \big( \xi_1 - \xi_2 -(1-10/\sqrt{3})\xi_3 - 6\xi_7\big)
                    \overline{m}^0_s
         + \frac{1}{4}\big( 2\xi_1 - 15\xi_2 - 3\xi_6 \big)
                      \big(6+ 5\overline{m}^0_s \big)
\\
&&       + 6\big( \xi_8 - 2\xi_5 \big) \bigg[
           4R \bigg( 1 + \frac{3\overline{m}^0_s m^2}{4\pi^2 F^2_0} \bigg)
         - \frac{m^2_K}{m^2} \bigg ] \bigg \}.
\end{eqnarray*}

    Because all dominating $UV$-divergent parts of meson-loop diagrams for
$K \rightarrow 2\pi \gamma$ decays are completely absorbed by their
$IB$-parts (see, for example, the corresponding remark in ref.
\cite{K2pigam-our}) the residual contributions of finite part of
one-loop diagrams to $DE$-amplitudes of $K \rightarrow 2\pi \gamma$
decays are small and can be neglected.
    The meson-loop corrections to $IB$-amplitudes of
$K \rightarrow 2\pi \gamma$ decays can be taken into account simply by using
the relations (\ref{IB}) and the results of one-loop calculations of
$K \rightarrow 2\pi $ amplitudes with $SP$-regularization \cite{DESY92-106}.

    The matrix element squared for $K^{\pm} \rightarrow \pi^{\pm} \pi^0\gamma$
decay, concerning with the distributions summed over both photon polarization,
can be presented as a sum
\begin{eqnarray}
|\,T(K^{\pm} \rightarrow \pi^{\pm} \pi^0 \gamma)\,|^2 =
                                            W_{IB}+W_{M1}+W_{E1}+W_{int},
\label{WKch}
\end{eqnarray}
    where
$$
W_{IB} = 3 \pi \alpha |A_2|^2
         \frac{  m^2_{\pi}(q\!\cdot\! k)^2 + m^2_K(q\!\cdot\! p_1)^2
              - (p_1\!\cdot\! k) (q\!\cdot\! p_1) (q\!\cdot\! k)}
             {(q\!\cdot\! p_1)^2 (q\!\cdot\! k)^2}
$$
    is the pure inner bremsstrahlung contribution;
\begin{eqnarray}
 W_{M1} &=& 4 \pi \alpha |h_{M1}|^2 \big[
           2(q\!\cdot\! p_1) (q\!\cdot\! p_2) (p_1\!\cdot\! p_2)
         - m^2_{\pi}\big( (q\!\cdot\! p_1)^2 + (q\!\cdot\! p_2)^2 \big) \big],
\nonumber
\\
 W_{E1} &=& \pi \alpha |h_{E1}|^2 \big[
           \big( (q\!\cdot\! (p_1 - p_2)\big)^2 (p_1 + p_2)^2
         + \big( (q\!\cdot\! (p_1 + p_2)\big)^2 (p_1 - p_2)^2 \big]
\label{WM1E1}
\end{eqnarray}
    are the contributions of magnetic and electric dipole transitions,
respectively;
$$
      W_{int} = 2 \sqrt{3} \pi \alpha \widetilde {R} \Big[
                 \re\Big( A_2 h^*_{E1} \Big)
                         cos \Big( \delta^1_1 - \delta^2_0 \Big)
               \pm \im\Big( A_2 h^*_{E1} \Big)
                         sin \Big( \delta^1_1 - \delta^2_0 \Big)
                                                        \Big],
$$
$$
\widetilde {R} =
                 \big( q \!\cdot\! (p_1 - p_2) \big) \bigg[
                 \frac{m^2_K - (q \!\cdot\! k)}{q \!\cdot\! k}
               - \frac{m^2_{\pi} + (p_1 \!\cdot\! p_2)}{q \!\cdot\!p_1}
                                                     \bigg]
               - \big( q \!\cdot\! (p_1 + p_2) \big) \bigg[
                 \frac{q \!\cdot\! (p_1 - p_2)}{q \!\cdot\! k}
               - \frac{m^2_{\pi} - (p_1 \!\cdot\! p_2)}{q \!\cdot\!p_1}
                                                      \bigg]
$$
    is the interference of the inner bremsstrahlung and electric dipole
amplitudes (since the photon polarization is not measured, the magnetic
dipole term does not interfere).

    Due to the charge conjugation properties (\ref{M1E1}) $E1$-transitions
contribute into the decay $K^0_1 \rightarrow 2 \pi \gamma$ of $CP$-even state
$K^0_1$ just as $M1$-transitions contribute into the decay
$K^0_2 \rightarrow 2 \pi \gamma$ of $CP$-odd state $K^0_2$.
    Thus,
\begin{eqnarray*}
  T\big( K^0_L \rightarrow 2 \pi \gamma \big) = T_{M1} + \varepsilon T_{E1},
\qquad
  T\big( K^0_S \rightarrow 2 \pi \gamma \big) = T_{E1} + \varepsilon T_{M1}.
\end{eqnarray*}
    The matrix element squared for
$K^0_{S,L} \rightarrow \pi^+ \pi^- \gamma$ decays can be presented
similarly to eq.(\ref{WKch}) as a sum of $IB$, $M1$, $E1$ and
interference contributions with $W_{M1}$ and $W_{E1}$ being determined
by the same expressions as (\ref{WM1E1}) where $h_{M1}$, $h_{E1}$
should be replaced to $h^{M1}_{S,L}$, $h^{E1}_{S,L}$ respectively, and
$$
W^{IB}_{K^0_S} = 4 \pi \alpha |T(K^0_S \rightarrow \pi^+ \pi^{-})|^2
         \frac{ m^2_{\pi}
               \big((q\!\cdot\!p_1)^2 + (q\!\cdot\!p_2)^2 \big)
               - 2(p_1\!\cdot\!p_2) (q\!\cdot\!p_1) (q\!\cdot\!p_2)}
              {(q\!\cdot\!p_1)^2 (q\!\cdot\!p_2)^2};
$$
$$
W^{int}_{K^0_S} = 4\pi \alpha \widetilde {R}^{'} \bigg[
               \frac{1}{\sqrt{3}} \re\Big( A_0^* h^{E1}_{K^0_S} \Big)
                         cos \Big( \delta^1_1 - \delta^0_0 \Big)
               + \frac{1}{\sqrt{6}} \re\Big( A_2^* h^{E1}_{K^0_S} \Big)
                         cos \Big( \delta^1_1 - \delta^2_0 \Big)
                                                        \bigg],
$$
$$
\widetilde{R}^{'} =
                 \big( q\!\cdot\!(p_1 - p_2) \big)
                 \big( m^2_{\pi} + (p_1\!\cdot\!p_2) \big)
                 \bigg(  \frac{1}{q\!\cdot\!p_2}
                       - \frac{1}{q\!\cdot\!p_1} \bigg)
               - \big( q\!\cdot\!(p_1 + p_2) \big)
                 \big( m^2_{\pi} - (p_1\!\cdot\!p_2) \big)
                 \bigg(  \frac{1}{q\!\cdot\!p_2}
                       + \frac{1}{q\!\cdot\!p_1} \bigg);
$$
$$
W^{IB}_{K^0_L} = \varepsilon^2 W^{IB}_{K^0_S}, \qquad
W^{int}_{K^0_L} = \varepsilon^2 W^{int}_{K^0_S}.
$$

    The results of the theoretical estimations of various
contributions to the branching ratios for the channels of
$K \rightarrow \pi \pi \gamma$ decay with the values of the parameters
$\xi_i$ taken from the nonleptonic kaon decay data
analysis (see eq.(\ref{ksi12345})\,) are presented in the Table
\ref{3}.
    The experimental cutoffs used in the experiments being taken
into account.
    The theoretical values of Table \ref{3} are in a good agreement
with the experimental data presented in the Table \ref{4}.
    For the estimates of the interference contributions, the
approximation $\Big( \delta^1_1 - \delta^2_0 \Big) \approx 10^o$ and
$\Big( \delta^1_1 - \delta^0_0 \Big) \approx - \delta^0_0$ can be used
with a good accuracy.
    The results of our previous calculations of the $s$-wave
$\pi \pi$-scattering phases and $K \rightarrow \pi \pi$ amplitudes
\cite{pl2,DESY92-106} were also involved in the numerical calculations.
    Fig.5 shows the spectrums of the center-of-mass $\gamma$
energy, $E_{\gamma}$, for $K^{+} \rightarrow \pi^{+} \pi^0\gamma$ and
$K^0_L \rightarrow \pi^+ \pi^- \gamma$ decays.

    It is worth noting that the decays
$K^{\pm} \rightarrow \pi^{\pm} \pi^0 \gamma$ were expected to be the most
suitable source of the experimental information about the direct
$CP$ violation in the charged kaon decays.
    Thus, for example, in \cite{K2pigamCP} the large values of charged
asymmetries of the branching ratio and differential distributions of
$K^{\pm} \rightarrow \pi^{\pm} \pi^0 \gamma$ decays were predicted.
    These large charged asymmetries were expected to originate from the
interference of the $IB$ amplitude and $E1$-transition.
    However, neither a charge asymmetry nor interference effects could
be found in experiments \cite{bolotov,smith,abrams}.
    This experimental situation may easily be understood on the basis
of the results for $K^+ \rightarrow \pi^+ \pi^0 \gamma$ decay
presented in Table \ref{3}.
    Indeed, the $DE$ amplitude is dominated by $M1$ transitions
arising from diagrams with anomalous vertices and the contribution of
$E1$ transition to the branching ratio of the decay
$K^+ \rightarrow \pi^+ \pi^0 \gamma$ is about of three orders of
magnitude smaller then $M1$ contribution.
    It is for this reason that interference between $E1$ and $IB$
amplitudes is suppressed and a charge $CP$-asymmetry was not observed
in the experiment.
    The value of charge asymmetry of branching ratio of
$K^{\pm} \rightarrow \pi^{\pm} \pi^0 \gamma$ decays, corresponding to
the imaginary part of the coefficient $c_5$ (\ref{Imc5}), is
$$
    |\Delta B(K^{\pm} \rightarrow \pi^{\pm} \pi^0 \gamma)| =
                1.52 \cdot 10^{-3} | \varepsilon' / \varepsilon |.
$$
    The dependence of charge asymmetries for branching ratio and the
spectrum of the center-of-mass $\gamma$ energy on top quark mass are
shown in fig.6.

%
\section*{Conclusions}
%

   In this paper the possibility of selfconsistent description of
nonleptonic and radiative kaon decays within chiral bosonized
Lagrangians have been demonstrated and $CP$-violation effects in
$K \rightarrow \pi \gamma^* \rightarrow \pi e^+e^-$ and
$K \rightarrow \pi \pi \gamma$ decays have been estimated with explicit
accounting of the gluonic and electromagnetic penguins.
   In particular, the $CP$-violating one-photon exchange mechanism of
$K^0_L \rightarrow \pi^0 e^+e^-$ decay seems to be available for the
experimental investigation and the direct $CP$-violating
$K^0_L \rightarrow \pi^0 e^+e^-$ transition can in fact compete with indirect
$CP$-violation contribution for $m_t \geq 100 GeV$.
   On the other hand it was shown that, within present limits of
experimental accuracy, effects of $CP$-violation cannot be observed in
the decays $K^{\pm} \rightarrow \pi^{\pm} \pi^0 \gamma$.

   The authors are grateful to G.Ecker for useful discussions and
helpful comments.
   One of the authors (A.A.Bel'kov) thanks the Institute of Elementary
Particle Physics, Humboldt-University, Berlin and DESY-Institute for
High Energy Physics, Zeuthen for their hospitality.
   He also grateful to DFG for support of investigations (Project Eb
139/1-1).
   Another of the authors (A.V.Lanyov) is grateful for the
hospitality extended to him at the DESY-Institute for High Energy
Physics, Zeuthen. We want to thank German Valencia for helpful
comments.

%
%

  %

\newpage
\begin{center}
                \Large Figure Captions \large
\end{center}

  Fig.1. Diagrams describing the one-photon exchange (a) and
two-photon intermediate state (b) contributions to the decays
$K \rightarrow \pi e^+ e^-$.\\

  Fig.2. Born diagrams of $K \rightarrow \pi \gamma^*$ transitions
(a-c) and meson loops (d) contributing to the terms in the amplitudes
proportional to $q^2(p+k)_{\mu}$ ($s$ -- strong vertices, $w$ -- weak
vertices, $ew$ -- electromagnetic-weak vertices, $em$ -- electromagnetic
vertices).\\

  Fig.3. Electromagnetic penguin diagram generating the additional
quark-leptonic operators $\O_{7,8}^{'}$.\\

  Fig.4. Born diagrams describing $K \rightarrow \pi \pi \gamma$
decays ($w$ -- weak vertices, $WZ$ -- anomalous Wess-Zumino vertices).\\

  Fig.5. Spectrum of the center-of-mass $\gamma$ energy for
$K^+ \rightarrow \pi^+ \pi^0 \gamma$ and
$K^0_L \rightarrow \pi^+ \pi^- \gamma$ decays. Dashed line is the $M1$
$DE$ contribution, dotted line is $IB$ contribution, dash-dotted line
is the contribution of the interference of $IB$ and $E1$ $DE$
amplitudes, and the solid line is their sum. For $K^0_L \rightarrow
\pi^+ \pi^- \gamma$ decay the interference of $IB$ and $E1$ $DE$ amplitudes
gives a negative contribution, and its absolute value is presented here.\\

  Fig.6. Dependence of the asymmetries of the branching ratio and
of the spectrum of the center-of-mass $\gamma$ energy for
$K^{\pm} \rightarrow \pi^{\pm} \pi^0 \gamma$ decays on the top quark
mass.\\

\newpage

\begin{table}[h]
\begin{center}
\caption{\label{1} Born and one-loop contributions to the amplitudes of
                   $K \rightarrow \pi \gamma^*$ transitions.}
\bigskip
\begin{tabular}{|c|c|c|c|c|}\hline
\multicolumn{5}{|c|}{\rule[-3mm]{0mm}{8mm}$K^+ \rightarrow \pi^+
\gamma^*$}\\ \hline
$E_{\pi}$(MeV)&$f^{(Born)}$&$\re f^{(loop)}$&$\im f^{(loop)}$&$|f^{(tot)}|$
     \\ \hline
141 &0.0276&-0.0129&-0.0021&0.0149 \\
166 &0.0264&-0.0126&-0.0009&0.0138 \\
192 &0.0251&-0.0114&   0   &0.0137 \\
217 &0.0239&-0.0105&   0   &0.0133 \\
242 &0.0226&-0.0100&   0   &0.0127 \\
265 &0.0215&-0.0096&   0   &0.0119 \\ \hline
\end{tabular}

\vspace*{1cm}

\begin{tabular}{|c|c|c|c|c|}\hline
\multicolumn{5}{|c|}{\rule[-3mm]{0mm}{8mm}$K^0 \rightarrow \pi^0
\gamma^*$}\\ \hline
$E_{\pi}$(MeV)&$f^{(Born)}$&$\re f^{(loop)}$&$\im f^{(loop)}$&$|f^{(tot)}|$
    \\ \hline
136 &0.00637&0.00141&0.000021&0.00778 \\
163 &0.00637&0.00137&0.000012&0.00775 \\
189 &0.00637&0.00133&0.000001&0.00771 \\
216 &0.00637&0.00130&    0   &0.00767 \\
242 &0.00637&0.00126&    0   &0.00764 \\
266 &0.00637&0.00122&    0   &0.00757 \\ \hline
\end{tabular}
\end{center}
\end{table}

%
\begin{table}[h]
\begin{center}
\caption{\label{2} The contributions of operators $\O_{5,8}$ and $\O'_7$
                   to the branching ratio of direct CP-violating
                   $K_L^0 \rightarrow \pi^0e^+e^-$ transition.}
\bigskip
\begin{tabular}{|r|c|c|c|} \hline
{\rule[-3mm]{0mm}{8mm}}$m_t$(GeV)&$B^{dir}_{\O_{5,8}}  \cdot
| \frac{\varepsilon'}{\varepsilon}|^{-2}$&$B^{dir}_{\O'_7} \cdot
| \frac{\varepsilon'}{\varepsilon}|^{-2}$&$B_{tot}^{dir}
\cdot | \frac{\varepsilon'}{\varepsilon} |^{-2}$ \\ \hline
 50 & $3.22 \cdot 10^{-8}$&$4.70 \cdot 10^{-7}$&$2.57 \cdot 10^{-7}$ \\
 75 & $9.72 \cdot 10^{-8}$&$1.24 \cdot 10^{-6}$&$6.47 \cdot 10^{-7}$ \\
100 & $2.73 \cdot 10^{-7}$&$2.36 \cdot 10^{-6}$&$1.05 \cdot 10^{-6}$ \\
125 & $7.39 \cdot 10^{-7}$&$3.93 \cdot 10^{-6}$&$1.38 \cdot 10^{-6}$ \\
150 & $1.89 \cdot 10^{-6}$&$6.19 \cdot 10^{-6}$&$1.57 \cdot 10^{-6}$ \\
175 & $4.59 \cdot 10^{-6}$&$9.50 \cdot 10^{-6}$&$1.60 \cdot 10^{-6}$ \\
200 & $1.09 \cdot 10^{-5}$&$1.49 \cdot 10^{-5}$&$1.64 \cdot 10^{-6}$ \\ \hline
\end{tabular}
\end{center}
\end{table}

%
\begin{table}[h]
\begin{center}
\caption{\label{3}Theoretical branching ratios of
                  $K \rightarrow \pi \pi \gamma$ decays}
\vspace*{1cm}
\begin{tabular}{|l|l|l|l|l|l|}
\hline
& & & & &\\
Decays & $B(IB)$ & $B(E1)$ & $B(M1)$ & $B(int)$ & $B(IB+DE)$\\
& & & & &\\
\hline
& & & & &\\
$K^{\pm} \rightarrow \pi^{\pm} \pi^0 \gamma$ &
$2.47 \cdot 10^{-4}$ & $8.74 \cdot 10^{-8}$ & $1.71 \cdot 10^{-5}$ &
$5.17 \cdot 10^{-6}$ & $2.70 \cdot 10^{-4}$\\
& & & & &\\
\hline
& & & & &\\
$K^0_S \rightarrow \pi^+ \pi^- \gamma$ &
$1.64 \cdot 10^{-3}$ & $4.88 \cdot 10^{-10}$ & $2.90 \cdot 10^{-13}$ &
$-3.32 \cdot 10^{-6}$ & $1.64 \cdot 10^{-3}$\\
& & & & &\\
\hline
& & & & &\\
$ K^0_L \rightarrow \pi^+ \pi^- \gamma$ &
$1.27 \cdot 10^{-5}$ & $1.54 \cdot 10^{-12}$ & $3.37 \cdot 10^{-5}$ &
$-1.23 \cdot 10^{-8}$ & $4.64 \cdot 10^{-5}$\\
& & & & & \\
\hline
\end{tabular}
\end{center}
\end{table}

%
\begin{table}[h]
\begin{center}
\caption{\label{4}Experimental branching ratios of
                  $K \rightarrow \pi \pi \gamma$ decays}
\vspace*{1cm}
\begin{tabular}{|l|l|l|p{4cm}|}
\hline
& & &\\
Decays & Experiment & $B(IB+DE)$ &  $B(DE)$\\
& & & \\
\hline
& Bolotov $\big( 87 \big)$ \cite{bolotov} &
          $ \big( 2.71 \pm 0.45 \big)\cdot 10^{-4}$ &
          $ \Big( 2.05 \pm 0.46 ^{+0.39}_{-0.23} \Big)\cdot 10^{-5}$ \\
& & & \\
$K^{\pm} \rightarrow \pi^{\pm} \pi^0 \gamma$
& Smith $\big( 76 \big)$ \cite{smith} &
        $\big( 2.87 \pm 0.32 \big ) \cdot 10^{-4}$ &
        $ \big (2.30 \pm 3.2 \big ) \cdot 10^{-5}$ \\
& & &\\
& Abrams $\big( 72 \big) \cite{abrams} $ &
         $ \big( 2.71 \pm 0.19 \big) \cdot 10^{-4}$ &
         $ \big(1.56 \pm 0.35 \pm 0.5 \big) \cdot 10^{-5} $ \\ [0.5ex]
\hline
& Taureg $\big( 76 \big) \cite{taureg}$ &
         $ \big( 1.84 \pm 0.10 \big) \cdot 10^{-3}$ &
         $ < 0.11 \cdot 10^{-3}$ \\
& & & \\
$K^0_S \rightarrow \pi^+ \pi^- \gamma$
& Burgun $\big( 73 \big) \cite{burgun}$ &
         $ \big( 1.9 \pm 0.4 \big)\cdot10^{-3}$ &
         $< 0.57 \cdot 10^{-3}$ \\
& & &\\
& Ramberg $\big( 92 \big) \cite{ramberg} $ &
          $ \big( 1.76 \pm 0.06 \big)\cdot 10^{-3}$ &
\\ [0.5ex]
\hline
& Caroll $\big(80 \big) \cite{carroll} $ &
         $ \big( 4.41 \pm 0.32 \big) \cdot 10^{-5}$ &
         $ \big( 2.89 \pm 0.28 \big) \cdot 10^{-5}$ \\
$ K^0_L \rightarrow \pi^+ \pi^- \gamma$ & & & \\
& Ramberg $\big( 92 \big) \cite{ramberg} $ &
          $ \big( 4.66 \pm 0.15 \big)\cdot 10^{-5}$ &
          $ \big( 3.19 \pm 0.16 \big)\cdot 10^{-5}$
\\ [0.5ex]
\hline
\end{tabular}
\end{center}
\end{table}

\end{document}